\documentclass[a4paper,fleqn,usenatbib,onecolumn]{mnras}

\usepackage{newtxtext,newtxmath}

\usepackage{etoolbox}

\usepackage{xcolor}
\usepackage{amsmath}
\usepackage{graphicx}
\usepackage{psfrag}
\usepackage{upgreek}

\newcommand{\beq}{\begin{equation}}
\newcommand{\eeq}{\end{equation}}
\newcommand{\nn}{\nonumber}
\newcommand{\rmd}{\mathrm{d}}

\newcommand{\brac}[1]{\left({#1}\right)}
\newcommand{\pd}[2]{\frac{\partial{#1}}{\partial{#2}}}
\newcommand{\td}[2]{\frac{\rmd{#1}}{\rmd{#2}}}
\newcommand{\curl}{\nabla\times}
\renewcommand{\div}{\nabla\cdot}

\newcommand{\be}{{\boldsymbol e}}

\newcommand{\bB}{{\boldsymbol B}}

\newcommand{\bv}{{\boldsymbol v}}

\title[Meissner effect in NSs]{The Meissner effect in neutron stars}
\author[S. K. Lander]{S. K. Lander\thanks{samuel.lander@uea.ac.uk}\\ \\
         School of Engineering, Mathematics and Physics, University of East Anglia, Norwich, NR4 7TJ, U.K.}

\begin{document}

\pagerange{\pageref{firstpage}--\pageref{lastpage}} \pubyear{0000}
\maketitle

\label{firstpage}

\begin{abstract}\
We present the first model aimed at understanding how the Meissner
effect in a young neutron star affects its macroscopic magnetic
field. In this model, field expulsion occurs on a dynamical timescale,
and is realised through two processes that occur at the onset of
superconductivity: fluid motions causing the dragging of field lines,
followed by magnetic reconnection. Focussing on magnetic fields weaker
than the superconducting critical field, we show that complete Meissner
expulsion is but one of four possible generic
scenarios for the magnetic-field geometry, and can never expel
magnetic flux from the centre of the star. Reconnection causes the
release of up to $\sim 5\times 10^{46}\,\mathrm{erg}$ 
of energy at the onset of superconductivity, and is only possible for certain favourable
early-phase dynamics and for pre-condensation fields
$10^{12}\,\mathrm{G}\lesssim B\lesssim 5\times 10^{14}\,\mathrm{G}$. Fields weaker or stronger than
this are predicted to thread the whole star.
\end{abstract}


\maketitle


\section{Introduction}

The hallmark of superconductivity in a material is the Meissner(-Ochsenfeld) effect,
characterised by an expulsion of magnetic flux and occurring once the
temperature $T$ drops below some critical value $T_c$. The nature of the
expulsion depends on the strength $B$ of the magnetic field through the
material prior to the onset of superconductivity. For weak fields
there is a complete expulsion of flux from the interior of the
superconducting sample; for sufficiently strong fields
superconductivity is destroyed and the conductivity is normal; for
intermediate field strengths there is a partial expulsion of flux. The
nature of this latter state depends on the kind of superconductivity
that is operative, but generally is characterised by narrow
structures of concentrated flux that have returned to the normal
regime, surrounded by superconducting regions with zero magnetic field.

In the laboratory, the magnetic field is imposed on a sample
externally, and once cooled below $T_c$ the sample
sets up a supercurrent that screens the external field and ensures the sample harbours no internal
magnetic flux. But this effect is also important in astrophysics,
being operative in one class of magnetic stars: neutron stars. The extremely
high density of a neutron-star core leads to a correspondingly high critical
temperature for superconductivity, and efficient neutrino cooling allows the star to
drop below this value shortly after birth.

A few features of the physics of neutron stars indicate that any
magnetic-flux rearrangement occurring due to the transition to
superconductivity will differ from the Meissner effect familiar from
terrestrial physics. For
one, their magnetic field is internal to the star, produced by the
persistent electric currents within the stellar fluid. As the Meissner
effect is not intrinsically dissipative, it is less obvious how to
expel this field than in the laboratory setup. Secondly, $T_c$
varies considerably throughout the core and is highest close to the
crust-core boundary; the Meissner effect will therefore
proceed gradually, on the 
cooling timescale, and rather than expelling the core magnetic
field may instead trap some of it. Finally,
terrestrial superconductors are solid, and the electron fluid becomes
superconducting. A neutron star core is entirely fluid, and instead of
the low-mass electrons, it is the proton fluid which forms a
superconductor. Because the neutrons, with their lower critical
temperature, only become superfluid considerably later on, the core
may be treated as a single fluid at the onset of
superconductivity. We are
therefore in an MHD regime, with the additional restrictions that that places on
the fluid flow, and therefore on how Meissner expulsion may be
realised.

The aim of this paper is to explore how the Meissner effect operates
in this setting, and what the result is likely to be on the star's
large-scale magnetic field.

\section{The onset of superconductivity: terrestrial versus
  neutron-star conditions}

Superconductivity occurs when it becomes energetically favourable for
two fermions to become coupled into a Cooper pair,
notwithstanding the Coulomb repulsion between them. In terrestrial
materials it is the electrons in the medium that may form Cooper
pairs; in neutron-star cores, the proton fluid. The critical
temperature, below which pairing sets in, depends upon the
properties of the medium, and is well understood for low-temperature
superconductors -- generally defined, in the terrestrial case, as those for which
$T_c<77\,\mathrm{K}$, the boiling point of liquid nitrogen. For
neutron-star cores the typical value is rather higher than this, 
$T_c\sim 10^9\,{\mathrm K}$, due to their vastly higher densities. This
means that it is still appropriate to treat them with the standard
theory of low-temperature superconductivity, whose fundamentals are
covered in several textbooks (we frequently draw upon \citet{tinkham} here). For temperatures $T>T_c$
superconductivity is destroyed, and the medium behaves according to
the usual equations of classical electrodynamics. Analogous to $T_c$, there
also exists a critical field strength $H_c$: superconductivity can be
destroyed by increasing the magnetic-field strength $B$ beyond $H_c$, as well as
by heating it above $T_c$.

In a typical, small, sample of terrestrial superconductor, $T$, $B$,
and the mass density $\rho$ will be very close to
constant, and once cooled sufficiently, the onset of superconductivity
will occur effectively instantaneously and globally throughout the
sample. The Meissner effect sets in, with the superconducting
electrons forming a supercurrent that screens the interior of the
sample from the externally-imposed field, and in the simplest case of
low temperatures and weak magnetic fields $B<H_c$, the magnetic flux is
effectively transported out of the bulk of the sample, into a thin
boundary layer. Like the onset of superconductivity, the action of the
Meissner effect in expelling magnetic flux is effectively
instantaneous, leaving a steady-state solution where the expulsion has
been completed. Really, although the term `Meissner expulsion' has
connotations of the process of field rearrangement itself, there is no
`Meissner term' that can simply be inserted into Ohm's law and thence
into Faraday's equation to describe the evolution of the field during
this phase. Instead, the
expression `Meissner effect' is used
to mean the final state once field rearrangement is over, where the
free energy is minimised. This endpoint \emph{can} be readily
calculated, from the
London equation for a magnetic field
$\bB$ in equilibrium:
\beq\label{london}
\nabla^2\bB=\frac{\bB}{\lambda^2}.
\eeq
The solution to this is a magnetic field that drops exponentially from its external
value to zero inside the medium, over a
lengthscale $\lambda$ known as the penetration depth, which is in good
agreement with experimental studies.
Though the final Meissner state is simple, a literature review
indicates that the very brief phase of flux 
rearrangement prior to the realisation of this state is not a research priority
for the field of terrestrial superconductivity, neither for
experimentalists nor theorists. It has even been argued that the standard
theory of superconductivity is not actually able to explain the
dynamics leading to Meissner expulsion \citep{hirsch12}.

In contrast to the terrestrial case, the onset of superconductivity
and the Meissner effect are slow processes for a neutron star. 
$T_c$ depends on $\rho$, which in turn varies by a factor of
up to $\sim 10$ in a neutron-star core. Because the core evolves
into a roughly isothermal state before the first onset of
superconductivity, the variation in $T_c$ throughout the core directly
corresponds to variation in the time at which different layers become
superconducting. Neither the density profile of the core, nor the
critical temperature, are known to a high degree of certainty, as a
result of differing approaches to treat the relevant microphysics -- but the qualitative
details are quite robust \citep{sedr_clark}, and are as follows. The first thin shell of superconducting matter
forms in the outer core at some radius $\mathcal{R}$, not far in from the crust-core boundary,
some minutes after the star's birth. The shell becomes thicker, expanding both
inwards and outwards on the cooling timescale for the star; it reaches
the crust-core boundary quickly, but its inward progress is
slower. Here the evolution may differ from model to model, depending
on the equation of state and gap model, and in some cases the
superconducting region may still not have reached the centre after
$10^6$ yr \citep{ho17}; though for some of the models presented here, the inner
part of the $T_c$ profile may be irrelevant, as the inward progress of
the superconducting shell can be arrested by a core region of magnetic
field amplified to the critical field strength (see section
\ref{later_evol}). Note that the core neutrons will also undergo pairing,
forming a superfluid, but at a substantially lower critical
temperature of $\sim 5\times 10^8\,\mathrm{K}$ \citep{page11}, corresponding to a
stellar age of $\sim 300\,\mathrm{yr}$; we will therefore treat them
as a normal fluid throughout this paper.

We will use a single critical temperature and cooling model
here for simplicity, but clearly the analysis could be repeated for
any other equation of state and gap model, were there motivation to do
so. In particular, following \citet{ho12} we use an approximation to
the proton pairing gap of \citet{chen93}, parametrised in terms of the
particle number density, which can then be 
rewritten to give $T_c$ as a function of $\rho$ (multiplying the number density by the average nucleon
mass). Converting this into an expression for $T_c$ as a function of
radius requires us to specify a density profile for the star, i.e. an
equilibrium solution for a particular equation of state. An obvious
simple choice would be the analytic solution from the Lane-Emden
equation for an $N=1$ polytrope, for which the density profile is
\beq\label{sinc}
\rho(r)=\rho_c\mathrm{sinc}\,\xi\ \ ;\ \ \xi=\frac{\pi r}{R_*},
\eeq
where $\rho_c$ is the central density.
However, although we will employ this later for an analytic expression
for a magnetic-field equilibrium (for which, we will argue, it is quite accurate),
for the calculation of $T_c$ it results in a rather unrealistic 
profile. Instead, we find that a tweak to equation \eqref{sinc}
gives a profile very close to that of the SLy equation of state
\citep{DH01} for
densities above the crust-core boundary value
$\rho_{\mathrm{cc}}=1.4\times 10^{14}\,\mathrm{g\, cm^{-3}}$ (which is
the only part of the profile relevant here):
\beq
\rho(r)=\rho_c\sqrt{\mathrm{sinc}(1.06\xi)}.
\eeq
Of course this expression is of no significance in the context of the
Lane-Emden equation -- it is just a convenient closed-form expression
that emulates a realistic core equation of state. The critical
temperature also depends on the proton fraction $x_p$. Following \citet{GAL},
we assume a simple linear dependence on $\rho$:
\beq\label{x_p}
x_p=0.1\frac{\rho}{\rho_c}.
\eeq
Utilising these results, and with the parametrisation of \citet{ho12}, we are
able to produce a critical-temperature profile.

\begin{figure}
\begin{center}
\begin{minipage}[c]{\linewidth}
  \includegraphics[width=0.5\linewidth]{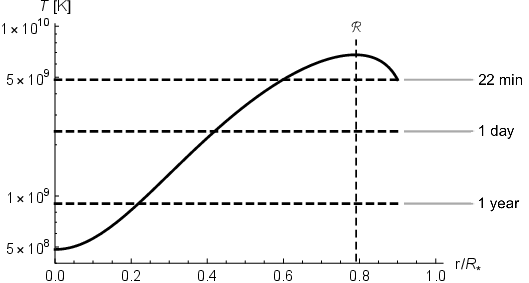}
  \hfill
  \includegraphics[width=0.45\linewidth]{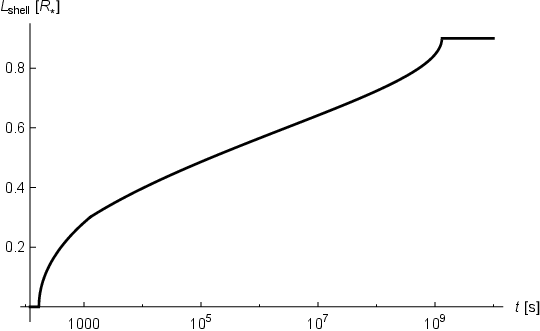}
\end{minipage}
\caption{\label{Tc_time}
Left: critical temperature for proton superconductivity, using an
approximate model similar to that of \citet{ho12}. Assuming an isothermal
core cooling according to the approximation of \citet{page06} and overlaying
these (horizontal) lines for different ages, we find that the first thin
shell of superconducting matter forms after $170$ seconds at the
dimensionless radius marked $\mathcal{R}$, and the
superconducting shell reaches the crust-core boundary after $22$
minutes. A few other ages are shown, and in the absence of magnetic
field the entire core will become superconducting after $42$ years. Right:
thickness of the superconducting shell, as a fraction of the stellar
radius $R_*$, as a function of time $t$.}
\end{center}
\end{figure}

Finally, to track the spreading of the superconducting region over
time as the core temperature drops, we use the closed-form cooling
prescription for an isothermal core of \citet{page06}, employing
values for the completely unpaired case (recalling that neutrons, representing
the bulk of the star's mass, are normal at this stage). Comparing this with the $T_c$ profile, we find
that the onset of superconductivity happens after $170$ seconds and at
a radius of $r/R_*=0.79$. The spreading of the superconducting region then
proceeds as shown in figure \ref{Tc_time}.

The aim of this paper is to consider possible scenarios by which the
Meissner effect is realised in a neutron star, over the long
phase during which the region of superconductivity spreads to
encompass most of the core. We will
argue that whether or not the magnetic field can be rearranged to
realise the Meissner effect depends on processes occurring just as a first
thin shell drops below $T_c$. The first process involves fluid
motions at the onset of superconductivity which, through Alfven's
frozen-flux theorem, drag field lines around; the second is the
subsequent process of magnetic reconnection that causes sufficiently
distorted field lines to pinch off and separate. If either process is
ineffective the Meissner state will not be realised, even though it is
the state minimising the free energy of the system for $B<H_c$. The field
configuration resulting from this first phase then dictates the later
evolution of the magnetic field in the face of a spreading region of superconductivity.

\section{Field rearrangement at the onset of superconductivity}

What happens when the first superconducting shell forms will dictate the
later magnetic-field structure, so this is a logical place to start
any analysis.

\subsection{Timescales}

The likely slow nature of Meissner expulsion from a superconducting
neutron-star core was discussed in the landmark paper of
\citet{BPP69a}, who identified the process with that of Ohmic decay:
\beq
\pd{\bB}{t}=-\frac{1}{4\pi}\curl\brac{\frac{c^2}{\sigma}\curl\bB}
\eeq
with characteristic timescale
\beq\label{tau_Ohm}
\tau_{\rm Ohm}=\frac{4\pi\sigma l_{\rm char}^2}{c^2},
\eeq
where $\sigma$ is the electrical conductivity and $l_{\rm char}$ some
characteristic lengthscale for the magnetic field. This implicitly assumes that the mechanism for transporting magnetic
flux in order to achieve Meissner expulsion must be a dissipative
one. But neutron-star core matter -- in its normal state -- forms an
outstandingly 
good electrical conductor, leading \citet{BPP69a} to conclude that
field expulsion would proceed on a timescale comparable with the age of the Universe. This assertion has
been reported uncritically ever since, but -- as we have already noted -- the condensation of the
core protons into a superconducting state is itself very slow and may
never be completed, so clearly one cannot make estimates for
the whole core in this manner. The spreading of the superconducting
region occurs on the cooling timescale, and is of great relevance, since it may be ongoing in
at least the younger observed neutron stars, and could
have important observational consequences. The only study to date that explores
this time-dependent process appears to be that of \citet{ho17}. These authors make a
significant new contribution to the problem: a detailed comparison of
the shortest magnetic-field evolutionary timescales (which they find
to be Ohmic decay and the drag due to scattering of electrons against
fluxtubes) with the cooling timescale. From this they find that the
cooling timescale is always shorter than that of field evolution, until a temperature of $T<10^8$ K
(equivalently, an age $\gtrsim 10^6$ yr), and therefore that in any
neutron star hotter (younger) than this, flux expulsion is not possible.

What these earlier studies and others all have in common is the
association of a
Meissner-type expulsion from a region of the star with decay of the
magnetic field there, but the Meissner effect is not itself 
instrinsically dissipative. In fact, as noted earlier it is not an evolutionary process
at all, but just a statement about the system's desired minimum-energy
equilibrium configuration (in its simplest form, it is found as a
solution to the time-independent London equation \eqref{london}).

Here we build on the arguments of \citet{ho17}, but argue that the
shortest relevant magnetic-field timescale is not due to either of the
secular dissipative effects they invoke, but rather the dynamical timescale
associated with the advection of magnetic field by the fluid
flow. This evolution is given by the expression
\beq
\pd{\bB}{t}=\curl(\bv\times\bB),
\eeq
where the velocity could, in principle, be approximately as high as
the Alfv\'en speed $v_A=B/\sqrt{4\pi\rho}$. Here we are interested in
the fastest possible evolution, so we set $|v|=v_A$, leading to a characteristic dynamical
timescale
\beq\label{tau_A}
\tau_A=\frac{l_{\mathrm{char}}}{v_A}=\frac{l_{\mathrm{char}}\sqrt{4\pi\rho}}{B}
=35\, l_{\mathrm{char},6}\rho_{14}^{1/2}B_{12}^{-1}\ \mathrm{sec},
\eeq
where we have employed the shorthand notation that a numerical
subscript $n$ on a variable means its value in cgs units divided by
$10^n$,
e.g. $l_{\mathrm{char},6}=l_{\mathrm{char}}/(10^6\,\mathrm{cm})$.
In equation \eqref{tau_A}, $\tau_A$ is vastly shorter than flux-dissipation timescales, but the
comparison with the cooling timescale $\tau_{\rm cool}$ is equivocal. Taking, for
example, the result from \citet{page06} for slow neutrino cooling
(i.e. indirect Urca processes) in the absence of any Cooper-pairing,
and using their fiducial values for heat capacity and
the temperature-luminosity relation, we have
\beq\label{tau_cool}
\tau_{\rm cool}\approx 3\times 10^6\, T_9^{-6}\ \mathrm{sec}.
\eeq
Evaluating this for
the peak critical temperature from before, $T_9=6.8$, gives
$\tau_{\rm cool}=30\,\mathrm{sec}$ -- identical (to the level of
approximation) to the prefactor from equation \eqref{tau_A} for $\tau_A$, and presaging our later
conclusions that the efficacy of the Meissner effect depends quite
sensitively on the strength and characteristic lengthscale of the
magnetic field at the onset of superconductivity.

We will explore this dynamical Meissner process in detail, regarding
it as one that achieves flux expulsion mostly through field rearrangement, rather
than a wholesale dissipation of flux.

\subsection{Restrictions}
\label{restrictions}

However the Meissner effect proceeds, it has to obey some basic
rules \citep{mestel_book}. Firstly, by integrating the Maxwell equation $\div\bB=0$ over the
volume enclosed by the initial superconducting shell at radius
$r=\mathcal{R}$ and converting it
to a surface integral over this shell, we see that the magnetic field
normal to the surface $B_\perp$ must be continuous across this surface:
\begin{align}
0 &= B_\perp^{\mathrm{in}}-B_\perp^{\mathrm{out}}\nn\\
  &= B_r^{\mathrm{in}}-B_r^{\mathrm{out}}\ 
    \textrm{ for a spherical surface }r=\mathcal{R}.
\label{BC_rule_1}
\end{align}
This result is universal, and so no evolution mechanism can simply
`cut' and rejoin radial field lines across the initial shell of superconductivity
in order to effect flux expulsion. That is, even in the
previously considered idea of flux expulsion via dissipation, it is
not enough for Ohmic decay simply to reduce the overall magnetic-field
strength across the superconducting shell -- it must qualitatively
rearrange it too.

Secondly, we define the magnetic flux through a surface $S$:
\begin{align}
\mathfrak{F}\equiv\int\bB\cdot{\bf d}{\boldsymbol S}
                        &= \int B_\perp\ dS\nn\\
              &= \int B_r\ \rmd S_r
  = \mathcal{R}^2\int B_r(\mathcal{R},\theta,\phi)\sin\theta\ \rmd\theta\rmd\phi
\ \ \textrm{for a spherical surface }r=\mathcal{R}.
\end{align}
In the ideal MHD limit, $\mathfrak{F}$ must be conserved during the advection of field lines by the fluid,
i.e.
\beq
\pd{\mathfrak{F}}{t}=0
 \implies \mathfrak{F}_{\rm in}=\mathfrak{F}_{\rm out}=\mathfrak{F}_0, 
\label{BC_rule_2}
 \eeq
 where $\mathfrak{F}_{\rm in},\mathfrak{F}_{\rm out},\mathfrak{F}_0$
 are, respectively, the magnetic flux across the inner and outer boundaries of the
 superconducting shell, and across the shell of radius $r=\mathcal{R}$ just
 prior to the onset of superconductivity.
This will be exactly satisfied for the dynamical-timescale process of dragging field lines
around by the fluid, and very well satisfied for almost all of the
epoch during which the superconducting shell expands, given that the cooling
timescale is almost certainly considerably shorter than the Ohmic
decay timescale. The one exception, where a significant change in
$\mathfrak{F}$ may occur on short timescales, is if there is a phase
of magnetic reconnection across the newly-formed superconducting
shell.

\subsection{Pre-condensation field configuration}

To gain any detailed understanding of how much the Meissner effect
acts, we need a quantitative model of the magnetic field $B_0$ immediately prior to the
first onset of superconductivity. Without this we cannot make any
predictions, since timescale and energy considerations both involve
the characteristic lengthscale and distribution of the magnetic
field. The restrictions \eqref{BC_rule_1}, \eqref{BC_rule_2} will also have
different implications for different field geometries. For example,
one could imagine a contrived initial field configuration that just so happens to have $B_\perp=0$ all
along the spherical shell at which superconductivity first sets in --
i.e. there are no field lines crossing the shell. Neither of the two
earlier restrictions applies, and as the
superconducting shell expands it is able just to push the outer zone
of the field further outwards, and the inner zone further inwards; it
can thus straightforwardly realise a complete Meissner expulsion. This
field configuration is, however, a pathological case: a poloidal field
would never naturally have such a geometry because, 
among other reasons, the radius $r=\mathcal{R}$ is of no physical significance to the
pre-condensed state of the star.

On the other hand, \emph{every} purely
toroidal field has the property of no field lines crossing
$r=\mathcal{R}$, potentially allowing for a straightforward expulsion from the growing superconducting
region (see later discussion, in section \ref{tor_mix_fields}). There
are, however, two strong arguments to dismiss toroidal fields as generic models for a
neutron-star field: firstly, all their field lines close within the
star, leaving $B=0$ outside and rendering the object undetectable as a
typical neutron star; secondly, such field configurations are
dynamically unstable \citep{tayler73} and so could not be long-lived.

In fact, the second argument against purely toroidal fields also
applies to purely poloidal fields: they suffer instabilities on
dynamical timescales and so will not be realised in nature
\citep{wright73}, meaning that any realistic field configuration must
contain both toroidal and poloidal components, whose exact form will not be
generic but rather determined by the birth dynamo that amplifies the
field, together with early-time dynamics and stability. However,
for the purposes of this paper the most important feature of any
realistic pre-condensation field configuration is that there will
always be field lines crossing the initial superconducting
shell. Now, without any straightforward route to a Meissner-expelled state,
the magnetic field evolution will involve the interplay between the
superconducting region wanting to realise its minimum-energy state, and the energy penalty
associated with stretching and breaking field lines in order to
realise such a state.

We wish to start with the simplest non-trivial, realistic,
pre-condensation field configuration, and one we can progressively
build upon to increase the level of realism. For this we will take an
axisymmetric, purely poloidal, dipolar, magnetic field. This case
embodies the basic and most important feature for our analysis: that
magnetic field lines will always cross the nascent superconducting
shell. If one understands this test case, then we argue later on
(section \ref{tor_mix_fields}) that more realistic, complex field
geometries -- featuring higher multipoles, a toroidal component, a
non-axisymmetric structure --- can be tackled incrementally, and do
not pose specific new conceptual challenges.

We will also assume that the magnetic field is weak enough so that the minimum
energy state is full flux expulsion (regardless of whether the star
can actually achieve this). Quantifying the strength of a `weak' field
warrants a brief digression. We first remark that different regions of
a neutron star may harbour protons with different superconducting
properties, known as type-I or type-II superconductors. We will not
discuss the differences in detail here (although see
e.g. \citet{tilley^2}, \citet{GAS}), since the classification is related to
the minimum-energy state of a superconductor in the presence of
stronger magnetic fields than those we consider here. We do, however, need
to understand the physical meanings of the four critical fields
related to these type of superconductivity: these are usually
denoted $H_{cI},H_c$ for type-I superconductivity and $H_{c1},H_{c2}$
for type-II superconductivity. Two of these fields represent upper limits: superconductivity is destroyed if 
$B>H_c$ ($H_{c2}$) for a type-I (type-II) superconductor, meaning the
stellar fluid then behaves as in normal MHD. The other two fields are
lower limits for flux penetration: if $B<H_{cI}$ ($H_{c1}$) for
a type-I (type-II) superconductor, the minimal-energy state
is full flux expulsion. Finally, magnetic fields with strengths
between these lower and upper limits will penetrate the
superconductor, but through small-scale flux structures rather than
the smooth local field distribution expected for normal MHD.

The critical
fields for a type-II superconductor depend only on quantities related
to the equation of state \citep{GAS}, in particular the proton fraction and
the effective proton mass\footnote{See e.g. \citet{chamel08} for profiles of
  these quantities.} and not the magnetic field itself. The upper critical
fields for type-I and type-II superconductors are also related in a
simple way, $H_{c2}=\sqrt{2}\lambda H_c/\tilde\xi$, where $\tilde\xi$ is the
proton coherence length. But the lower critical field of a type-I superconductor is
geometry-dependent -- for example, for a spherical superconductor in a
vertical field, $H_{cI}=(2/3)H_c$; in other geometries $H_{cI}$ may be
as high as $H_c$ or as low as zero.

The critical fields do not all vary in the same way with the star's density (equivalently,
radius), but ignoring the very specific case where $H_c$ drops to zero
for the type-I case, the variation in the lower critical fields $H_{cI}$ and $H_{c1}$ will
generally only be by a small factor, mainly in the rough range
$5\times 10^{14}-10^{15}\,\mathrm{G}$ (e.g. \citet{GAS}). Our priority is to establish a single,
representative, cut-off value below which a magnetic field will be
energetically favoured to be expelled from the 
superconducting region, without needing to change treatment above a
certain density, nor for a certain field configuration. Because the
variation in lower critical fields is relatively small, we will simply
assume a spatially constant critical field, and
re-appropriate the symbol $H_c$ to denote this value, remaining
agnostic about the type of superconductivity, i.e. by `weak fields' we
mean the case
\beq
B<H_c\ ,\ \textrm{where }H_c=\mathrm{const}=5\times 10^{14}\,\mathrm{G}.
\eeq
Although at least some neutron stars probably host interior fields
$B>H_c$, it will be easier to tackle that case once the weak-field
case is better understood. We briefly discuss stronger fields at the
end of this paper, but defer any detailed calculations to future
work.

\subsection{Stretching of field lines during the onset of
  superconductivity}
\label{field_stretch}

Having established a representative pre-condensation magnetic field
and explored restrictions on how this field can be rearranged, we are
now in a position to describe our scenario for realising a
Meissner state on a dynamical timescale. This scenario consists of two
steps: firstly, distorting the field through advection into a
geometry amenable to magnetic reconnection, and secondly, the
reconnection phase itself. We consider the first step in this
subsection, and the second in the following subsection. The first step
envisages a transitional phase between the normal and superconducting phases:
that, at the onset of superconductivity,
fluid motions (treated with normal MHD) seek to drag the magnetic field into the
lower-energy state desired by the superconducting shell. To start
with, then, we need to understand whether this is plausible.

The minimum thickness for which the incipient shell can display
its superconducting properties is given by the penetration depth $\lambda$,
which for the outer part of the neutron-star core (where
superconductivity will first develop) is given by \citep{mendell91}
\beq
\lambda\approx 10^{-11}\ \textrm{cm.} 
\eeq
Clearly the shell's thickness will expand to a huge multiple of
$\lambda$ almost instantaneously, well within a typical Alfv\'en
timescale, which by equation \eqref{tau_A} is of order seconds. This
suggests that long before any fluid motions can act to distort the
initial magnetic field, the shell will be firmly in the
superconducting phase, invalidating the whole approach of this paper. However, the
fact that the incipient superconducting shell is -- by definition --
at a temperature $T\approx T_c$ changes this conclusion.

The penetration depth $\lambda$ and critical field $H_c$ are
temperature-dependent, and at temperatures close to $T_c$ differ from
their zero-temperature values $\lambda_0,H_{c_0}$ in the following
manner \citep{tinkham}:
\beq
\frac{\lambda(T)}{\lambda_0}=\left[1-\brac{\frac{T}{T_c}}^4\right]^{-1/2}
\ ,\ 
\frac{H_c(T)}{H_{c_0}}=1-\brac{\frac{T}{T_c}}^2.
\eeq
Consider the state of the star 10 seconds\footnote{Chosen for the sake of
  definiteness; our conclusions do not rely on the specific time.} after the onset of
superconductivity. After this time the superconducting shell has
already expanded to $0.05R_*$, i.e. $6\times 10^4\,\mathrm{cm}$ for a
12-km star: likely to be far longer than the characteristic
lengthscale for any fluid motion (this is discussed more in section
\ref{res_rec}). The critical temperature at the inner and outer
surfaces of the shell is (by definition) exactly equal to
the current temperature of the core, whilst within the shell at
$r=0.79R_*$ -- the point of first onset of superconductivity -- the
temperature is $0.96\%$ below its critical value. This
corresponds, using the above relations, to a penetration depth $5.2$
times longer than $\lambda_0$ and a critical field $53$ times
lower than $H_{c_0}$. The former modification is clearly not
significant ($\lambda$ remains vastly shorter than any relevant
hydrodynamic lengthscale), but the latter is dramatic. Taking our
fiducial value of $H_{c_0}=5\times 10^{14}\,\mathrm{G}$, it would mean
that the $0.05R_*$-thick superconducting shell
would have a critical field no higher than
$10^{13}\,\mathrm{G}$ in its centre, and lower towards its boundaries.
Given that the shell need not be this thick before field rearrangement
occurs, $H_c(T)$ would be even lower and very likely weaker than the existing field in
that region; from this we conclude that it is reasonable to treat the
early-stage dynamics in the shell with normal MHD.

\begin{figure}
\begin{center}
\begin{minipage}[c]{0.3\linewidth}
\includegraphics[width=\linewidth]{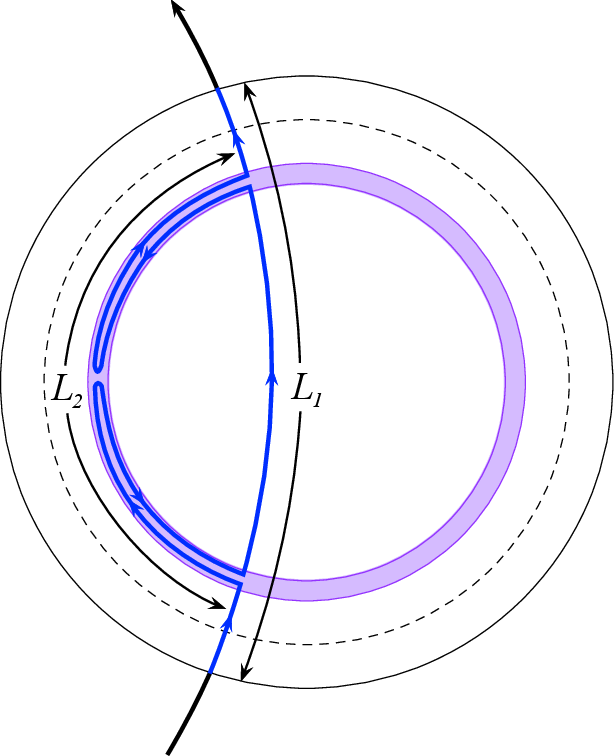}
\end{minipage}
\caption{\label{stretched_field_line}
Geometry of a field line distorted from its original length $L_1$ due
to shearing in the incipient superconducting shell (the thick purple circle).}
\end{center}
\end{figure}

For a simple dipolar geometry, field lines need to be advected a
distance comparable with the stellar radius, i.e. $\approx
10^6\,\mathrm{cm}$, in order to produce a geometry where reconnection
can later occur; see figure \ref{stretched_field_line}. If the Alfv\'en speed is too low, this advection will
not have time to produce such a geometry before the region is
enveloped by the superconducting region. To estimate the
characteristic \emph{angular} lengthscale $\mathcal{L}_\parallel$ associated with advection in the incipient
shell, we set the Alfv\'en \eqref{tau_A} and cooling \eqref{tau_cool} timescales equal to one another for
the appropriate values for the onset of superconductivity
($T_9=6.8,\rho_{14}=4.3$) within our model, yielding the result
\beq\label{L_par}
\mathcal{L}_{\parallel,6}\approx 1.3B_{12}.
\eeq
From this we conclude that hydromagnetic motions can 
rearrange the field in the nascent superconducting shell sufficiently
(i.e. over a $10^6$-cm scale) if the pre-condensation magnetic field strength
\beq\label{B_L_par}
B_0\gtrsim 10^{12}\,\mathrm{G}.
\eeq
Why should these motions set in to start with? Firstly, it
is natural to assume that the star is not strictly static; even
motions from the star's birth may not have been entirely dissipated by
viscosity in that time. Secondly, even if the
star had managed to achieve a strict equilibrium with the
pre-condensation field, the change of magnetic force from the
normal-matter Lorentz force to the corresponding magnetic force for a
superconductor would itself be enough to violate the equilibrium and
hence induce fluid motion.
Thirdly, the condensation energy density
\beq
\frac{H_c^2}{8\pi}=f_n^0-f_s^0
\eeq
represents the difference in Helmholtz free energies at $B=0$ for the
superconducting $f_s^0$ and normal $f_n^0$ states\footnote{This
  expression is for the condensation energy at zero temperature; the
  energy will be substantially smaller close to $T_c$.}, so that the
Meissner state is able to minimise free energy, even though it will involve distorting
field lines away from the pre-condensation state, and thus increasing
the overall magnetic energy \citep{annett}.

The detailed dynamics will inevitably
be complex and are beyond the scope of this first analysis, and we will
just assume that they act to drive the magnetic field in the
superconducting region towards its minimum-energy state of full
expulsion. We come back to this point in the Discussion.

Fluid motions that stretch the field
lines also increase the total magnetic energy of the configuration,
and for full Meissner expulsion the fluid would need to advect the
field lines in such a way as to bunch them all up in a very small
volume of the superconducting shell, where they are highly distorted,
have a small local characteristic lengthscale, and hence are
susceptible to reconnection. Reconnection pinches the field lines off
across the shell and allows for complete Meissner expulsion, as
discussed in the following subsection; for this process we need to invoke
Ohmic decay. But first we consider the case of gradually distorting
the magnetic field lines, when the characteristic lengthscale is still
too long to need to worry about dissipative effects, and we can assume
infinite conductivity of the matter, with the field lines
perfectly frozen into the fluid.

Consider an infinitessimally thin fluxtube of cross-sectional area
$dA$ running along the entire interior length $L_1$ of a magnetic
field line, before the onset of superconductivity. It is likely safe to assume, to leading order, that any
internal distortion of the field line will have no effect on the part
of the field line that extends outside the star, and so we may ignore the
exterior field throughout this calculation. We define the magnetic energy of the fluxtube
$\mathfrak{E}_{\rm mag}$ as the (field) line integral of the magnetic-energy
density $B^2/8\pi$, multiplied by $\rmd A$:
\beq\label{emag_ft_orig}
\mathfrak{E}_{\rm mag}=\rmd A\int_{L_1} \frac{[B(s)]^2}{8\pi}\ \rmd s,
\eeq
where $s$ is a parameter defining distance along a field line.

Now assume that the action of fluid motions causes maximal distortion
of the (still single) field line, dragging it towards the equator
within the superconducting shell, as shown in figure \ref{stretched_field_line}.
When the field line is thus distorted,
it gains an additional length $2L_2$ -- one factor of $L_2$ each for
its journey towards the equator, and back again. Denoting the radius
at which the superconducting shell first appears by $\mathcal{R}$, the
length $L_2$ must be in the interval
\beq
0\leq L_2\leq \pi\mathcal{R},
\eeq
and the total length after reconnection is the sum of the lengths of the now-separated two
field lines, i.e. $L_1+2L_2$.
The fluxtube energy also gains an additional piece from the stretching,
with the total now being related to the sum over the line integrals
along both $L_1$ and $L_2$:
\beq\label{emag_ft_stretch}
\mathfrak{E}_{\rm mag}
  =\rmd A\brac{\int_{L_1} \frac{[B(s_1)]^2}{8\pi}\ \rmd s_1
                         + 2\int_{L_2} \frac{[B(s_2)]^2}{8\pi}\ \rmd s_2},
\eeq
where the two parametrising functions $s_1,s_2$ may be different.

This increase in magnetic energy must be sourced from somewhere, be it
kinetic energy from residual fluid motion remaining from the birth
phase, the condensation energy, or any other physics related to the
onset of superconductivity. It is certainly unlikely
that the star will be entirely static, so there may be a substantial well
of kinetic energy to draw upon, but there is no obvious way to
quantify this whilst retaining the generality of the model. Instead,
as an indicative upper limit on how much the field can be distorted,
we will regard the magnetic energy increase as being sourced solely by the
condensation energy in the relevant superconducting shell. Calculating
this is problematic, since (unlike terrestrial calculations) it is not
obvious what volume of shell to use, and because within this shell
$H_c(T)$ could be substantially lower than the zero-temperature value
we mean when we write $H_c$. To give ourselves an order-of-magnitude
idea of the kind of well of energy we may be able to draw on, we
estimate the condensation energy using the zero-temperature
$H_c=5\times 10^{14}\,\mathrm{G}$ and a shell extending from $0.8-0.9R_*$:
\beq\label{E_cond}
\int_{\rm shell} \frac{H_c^2}{8\pi}\,\rmd V
\approx  \frac{4\pi [(0.9R_*)^3-(0.8R_*)^3]}{3}\frac{H_c^2}{8\pi}
\sim 10^{46}\,\rm{erg},
\eeq
assuming a stellar radius $R_*=12\,\rm{km}$. 

This suggests -- somewhat counterintuitively -- that Meissner
expulsion from the superconducting region actually \emph{increases}
the total magnetic energy of the star, at least in the initial
field-line-stretching phase; we can get a qualitative estimate of this
increase by assuming the stretching roughly doubles the average length
of field lines (based on the sketch of figure
\ref{stretched_field_line}), and therefore also roughly doubles the
magnetic energy. Some, and possibly most, of this
additional energy will be lost if there is a reconnection phase
afterwards; we will make quantitative calculations of these values for
a specific model in the following section.

Finally, we recall that Meissner expulsion represents the system
seeking a lowest-energy state where $B=0$, and so is not expected
unless $B\lesssim H_c$. Essentially the same physics is also 
encapsulated by the requirement that any field-line stretching
increases the magnetic energy by at most the estimate of equation \eqref{E_cond}.

\subsection{Resistivity and reconnection}
\label{res_rec}

We assume that the main cause of flux rearrangement during the onset
of superconductivity is due to advection by fluid motions, with the
field being perfectly frozen into the fluid, as is the case where the
conductivity is infinite. Through this process, the field lines in the
thin initial shell of superconducting matter could be dragged around
significantly; see figure \ref{stretched_field_line} and the middle row of cartoons from figure
\ref{flux_scenarios}. However, Meissner expulsion can not be 
completed unless the highly sheared magnetic field lines ultimately
reconnect. We assume the matter can still be treated as in a normal
state here, since reconnection should occur (\emph{if} it
occurs) at the onset of superconductivity, when $T\approx T_c$ and $B>H_c(T)$, as
discussed in the previous subsection. Even if there is significant
supercurrent screening, however, causing a substantial reduction in
the value of $B$ compared with elsewhere in the star, this will not
affect the calculation, since neither reconnection nor cooling depend
(to leading order) on the magnetic-field strength itself.

Whether the advected field discussed in the previous section can
reconnect depends on whether or not this process is faster than the
spreading of the superconducting region as the star cools, i.e. the
efficacy of reconnection in this scenario can be measured with the ratio:
\beq
\frac{\tau_{\mathrm{rec}}}{\tau_{\mathrm{cool}}}.
\eeq
The smaller the value of this ratio, the more efficiently reconnection
should be able to alter the geometry of the distorted field in the
thin shell where superconductivity begins.
Amongst the various mechanisms for magnetic reconnection in the
literature, there are essentially two kinds of relevance here; both are related to Ohmic decay (and so depend
upon the resistivity of neutron-star
matter in its normal state), but they differ in the assumed
characteristic magnetic-field lengthscale and small-scale
magnetic-field structure. The first is the classic Sweet-Parker
reconnection \citep{parker57,sweet58}, which acts on the macroscopic field; the
second is stochastic reconnection, frequently invoked to explain the phenomenon
of fast dynamo action in stars \citep{LV99}, where jagged microscopic field
structures allow for a faster diffusive rearrangement of magnetic field. 

We begin with the Sweet-Parker mechanism, whose reconnection timescale
is simply that of Ohmic decay, $\tau_{\rm rec}=\tau_{\rm Ohm}$. To
evaluate this we adopt the density- and temperature-dependent
expression for $\sigma$ from \citet{BPP69b}, together with our previous approximation \eqref{x_p} for proton
fraction, which gives us
\beq\label{sigma}
\sigma=4.7\times 10^{26}\,\rho_{c,15}^{-3/2}\rho_{15}^3 T_{10}^{-2}\,\mathrm{s^{-1}}.
\eeq
Now plugging this into equation \eqref{tau_Ohm} and using the cooling prescription of
\citet{page06}, we arrive at the ratio
\beq\label{tau_ratio_SP}
\frac{\tau_{\mathrm{rec}}}{\tau_{\mathrm{cool}}}
= 2.2\times 10^6\rho_{c,15}^{-3/2}\rho_{15}^3 T_{10}^4
   l_{\rm char,1}^2.
\eeq
The large prefactor suggests that Sweet-Parker reconnection is
unlikely to be effective at flux rearrangement at the onset of superconductivity, given the plausible
range in which the other quantities can vary. For our particular
model, the onset of superconductivity occurs at
$T=6.8\times 10^9\,\textrm{K}$ and at a radius $0.79R_*$, which
corresponds to a density $\rho_{15}=0.43$ for a model with central
density $\rho_{c,15}=1$. Let us insert these values into equation
\eqref{tau_ratio_SP}, and also change notation from $l_\mathrm{char}$
to $\mathcal{L}_\perp$, to emphasise the fact that the relevant
lengthscale is related to the component of the magnetic field/fluid
motion normal to the nascent shell of superconductivity (cf equation
\eqref{L_par} for $\mathcal{L}_\parallel$):
\beq
\frac{\tau_{\mathrm{rec}}}{\tau_{\mathrm{cool}}}
= 3.7\times 10^4 \mathcal{L}_{\perp,1}^2,
\eeq
i.e. that only a characteristic field lengthscale $\mathcal{L}_\perp\lesssim 10^{-3}\,\textrm{cm}$
would allow for effective reconnection. A large-scale
configuration like a dipolar field, with $\mathcal{L}_\perp\sim 10^6\,\textrm{cm}$,
will therefore clearly not experience any qualitative rearrangement on
the onset of superconductivity. If field lines are, however, strongly
distorted in the initial shell of superconductivity, the lengthscale
becomes shorter. Determining this lengthscale quantitatively requires
a full, numerical, solution of the MHD equations for this scenario,
but we can at least determine a minimum characteristic
lengthscale. This will depend on whether the main dissipative
mechanism acting on the MHD flow is due to 
resistivity $c^2/(4\pi\sigma)$ or viscosity $\upnu$.

Although
the neutron star is only $170\,\rm{s}$ old, it is still comfortably in
the neutrino-transparent phase \citep{BL86}, for which viscosity is dominated by
the contribution from neutron-neutron scattering \citep{flowers_itoh},
with \citep{CL87}
\beq\label{nu}
\upnu\approx 19\,\rho_{15}^{5/4}T_{10}^{-2}\mathrm{cm^2\, s^{-1}}.
\eeq
At the onset of superconductivity in our model, this gives a value of
$\upnu\approx 10\,\mathrm{cm^2\, s^{-1}}$. We can now combine
equations \eqref{sigma} and \eqref{nu} to yield an
expression for the magnetic Prandtl number $\mathrm{Pm}$, the ratio of
resistivity to viscosity:
\beq
\mathrm{Pm}=1.2\times 10^8\, \rho_{c,15}^{-3/2}\rho_{15}^{17/4} T_{10}^{-4}.
\eeq
At the onset of superconductivity this gives $\mathrm{Pm}=1.6\times
10^7$, and shows that the shortest
characteristic lengthscale will be set by the viscosity, which acts on
a fluid flow that can be no faster than the Alfv\'en speed
$v_A=B/\sqrt{4\pi\rho}$; now combining $\upnu$ and $v_A$ and with suitable
parametrisations, we arrive at a lower limit for the characteristic
lengthscale of the viscous MHD flow:
\beq
l_{\textrm{visc}}=\frac{\upnu\sqrt{4\pi\rho}}{B}
= 2.1\times 10^{-3}\, \rho_{15}^{7/4} B_{12}^{-1}T_{10}^{-2}\,\textrm{cm}.
\eeq
Inserting values for the onset of superconductivity, within the model
we adopt, yields a viscous scale of
\beq\label{l_visc}
l_{\mathrm{visc}}=1.1\times 10^{-3}B_{12}^{-1}\,\mathrm{cm}.
\eeq
Reconnection cannot proceed if it requires a shorter lengthscale than
$l_{\rm visc}$, so we compare the latter with the $l_{\rm char}$
required to produce a ratio of unity in equation \eqref{tau_ratio_SP}:
this shows that for Sweet-Parker reconnection to be at least
marginally effective, it requires
\beq\label{B_L_perp}
B_0\gtrsim 10^{12}\,\mathrm{G}.
\eeq
This only proves that
this reconnection scenario is not \emph{implausible} for a
maximally-distorted pre-condensation field $\gtrsim 10^{12}\,\mathrm{G}$; only
simulations can determine whether it is actually likely. Note that
Meissner expulsion through the mechanism discussed in this paper
requires both significant advection of the fluid flow and effective
reconnection, and these lead to identical minimum-field requirements,
equations \eqref{B_L_par} and \eqref{B_L_perp} respectively. This is
just a coincidence, but it does make the lower-field bound more robust.

Next we examine the mechanism of stochastic reconnection. This
mechanism has proved invaluable in understanding the problem of fast
dynamos in astrophysics, i.e. dynamos whose field-amplification rate
becomes independent of resistivity in the limit of vanishing
resistivity. The original work on this topic gives a reconnection speed
$v_{\mathrm{rec}}$ that depends upon the rms velocity due to energy
injection at the stochastic scale $v_T$, and upon the ratio of two key
lengthscales, the width of the reconnection region $L_x$ and the scale
of energy injection $l_\mathrm{inj}$ \citep{LV99}:
\beq
v_{\mathrm{rec}}\lesssim v_T\min\left\{\sqrt{\frac{L_x}{l_\mathrm{inj}}}, \sqrt{\frac{l_\mathrm{inj}}{L_x}}\right\},
\eeq
with (as required by the motivation of understanding fast dynamos) no dependence on $\sigma$. This
expression depends on details of the energy injection, but an
indicative range of values for the reconnection speed is
$v_{\mathrm{rec}}=(0.01-0.1)v_A$ \citep{kowal09}.

The above result is valid when viscosity is weak, which encompasses
most astrophysical fluid settings where dynamos
and field rearrangement are expected, but we just saw that neutron
stars are an exception to this, with $\mathrm{Pm}\gg 1$. In this
large-$\mathrm{Pm}$ regime the previous
reconnection speed is modified to \citep{jafari18}:
\beq
v_{\mathrm{rec}}\lesssim v_T\frac{L_x}{l_\parallel}
                                       \frac{\mathrm{Re}^{1/4}\mathrm{Pm}^{-1/2}}{1+\ln(\mathrm{Pm})},
\eeq
where $l_\parallel$ is a parallel eddy lengthscale.                                       
Note that in this case reconnection is no longer truly `fast', as it regains
an implicit dependence on resistivity through the $\mathrm{Pm}$
terms. The two lengthscales in this relation are not likely to be
independent of one another, as both relate to the field structure produced by the
MHD flow and therefore should scale with (or be equal to) the
viscous scale; we will therefore just pessimistically assume the ratio $L_x/l_\parallel$ to be unity,
though it could be substantially larger. For the rms velocity we
assume $v_T=0.01v_A$, and we calculate the Reynolds number
based on a flow moving at $v_A$, giving $\mathrm{Re}=1400\, B_{12} l_{\mathrm{char},1}$.
Then, evaluating the ratio in
the above expression, we arrive at a stochastic reconnection speed
(reduced by viscosity) of
\beq
v_{\mathrm{rec}}=0.012\,B_{12}^{5/4}l_{\mathrm{char},1}^{1/4}\mathrm{cm\, s^{-1}}.
\eeq
The ratio of reconnection to cooling timescales in this case, for our
fiducial model parameters for the onset of superconductivity, is
therefore
\beq
\frac{\tau_{\mathrm{rec}}}{\tau_{\mathrm{cool}}}
=3B_{12}^{-5/4}l_{\mathrm{char},1}^{3/4}.
\eeq
Although this has a far smaller pre-factor than for Sweet-Parker
reconnection, equation \eqref{tau_ratio_SP}, this is counteracted by
the weaker dependence on $l_{\rm char}$, and setting the above ratio
equal to unity indicates that the stochastic reconnection mechanism
is effective on somewhat less fine magnetic field structures than
Sweet-Parker reconnection,
$l_{\mathrm{char}}\lesssim 0.2B_{12}^{5/3}\,{\mathrm{cm}}$. This
lengthscale is also more comfortably above the viscous scale, equation
\eqref{l_visc}, at a field strength $B_{12}=1$. This calculation could therefore
be invoked to push the lower limit for partial/complete Meissner expulsion
below $10^{12}\,\rm{G}$ -- however, given the crude
assumptions we have made about parameters related to stochastic reconnection, and the fact that the work of
\citet{jafari18} was doubtless never intended for the kinds of extreme
magnetic Prandtl numbers we consider here, we find it safer to retain
$10^{12}\,\rm{G}$ as an indicative lower limit.

To summarise, understanding the role of reconnection at the onset of
superconductivity is hampered by the complexity of the process, even
with the simplification that it occurs in normally-conducting matter. Sweet-Parker
reconnection involves one characteristic lengthscale; stochastic
reconnection involves at least two -- and none of these can be
satisfactorily analysed without simulations. We can however draw some
useful conclusions. MHD flows in the high-$\mathrm{Pm}$ regime
expected in a young neutron star are capable of producing field
structures on a lengthscale where reconnection can take
place. Furthermore, if the
average field strength $B\gtrsim 10^{12}\,\mathrm{G}$ then
reconnection is faster than cooling; the field may be rearranged
before the superconducting shell spreads too far. Weaker fields will
not be substantially rearranged; for stronger fields the efficacy of
reconnection increases roughly linearly with field strength.

\subsection{Four limiting cases}

\begin{figure*}
\begin{center}
\begin{minipage}[c]{1.0\linewidth}
\includegraphics[width=\linewidth]{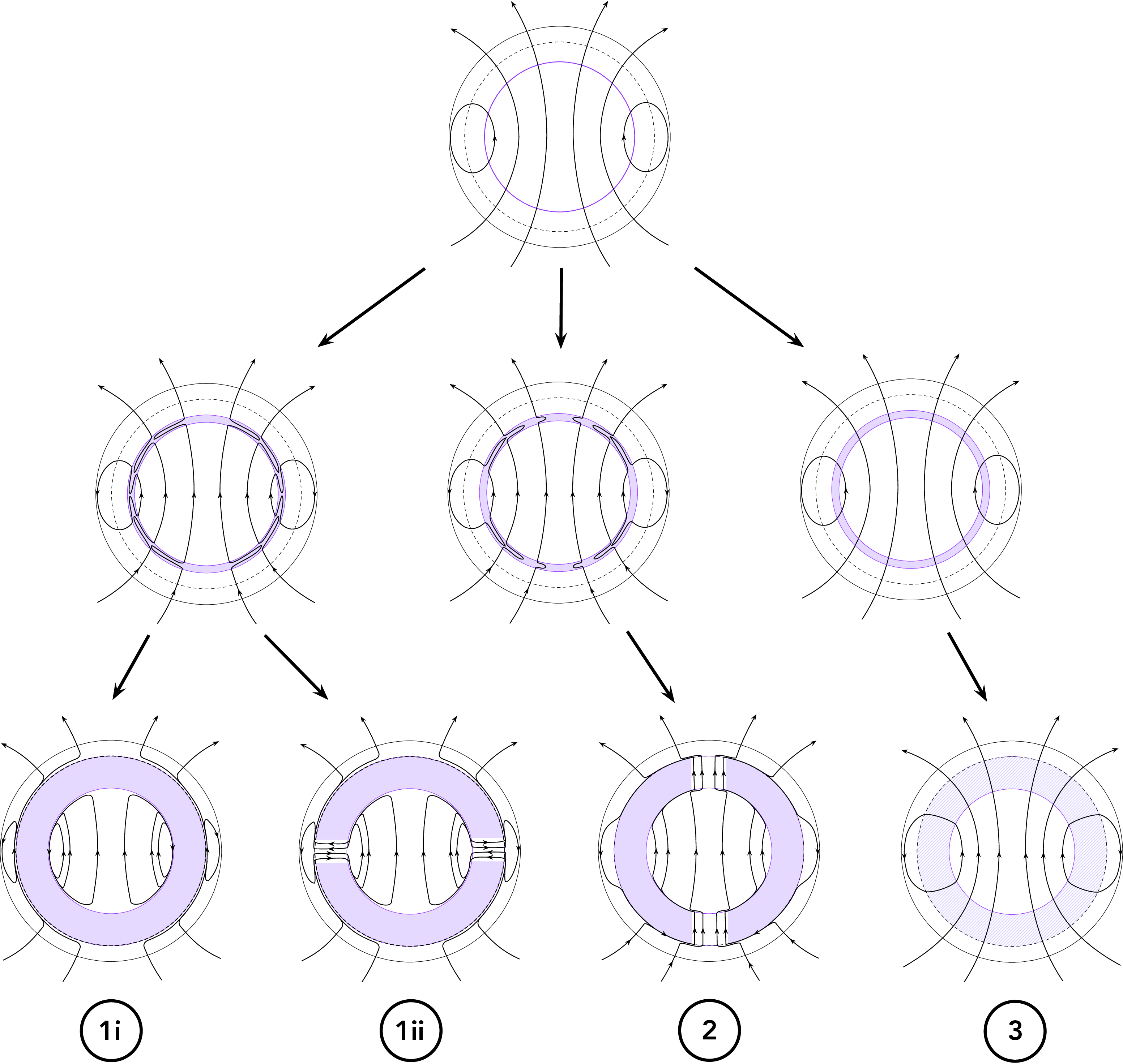}
\end{minipage}
\caption{\label{flux_scenarios}
 Four different scenarios for the rearrangement of flux due to the
expanding region of superconductivity, all for $B<H_c$. An initially
dipolar poloidal magnetic field (top row) may be advected by fluid motions
(middle row) and then experience reconnection (bottom row). The
(in)efficacy of these processes leads to four post-condensation field
configurations.}
\end{center}
\end{figure*}

From the previous two subsections, we saw that the range of field
strengths for which a partial or complete Meissner expulsion
\emph{could} occur is limited to
\beq
10^{12}\,\mathrm{G}\lesssim B_0\lesssim 5\times 10^{14}\,\mathrm{G},
\eeq
but that even within this range, expulsion depends on the fluid flow and reconnection
properties of the star at the point of formation of the incipient
superconducting shell. It is therefore now instructive to
consider four limiting cases of how the magnetic field may be affected,
and to examine the circumstances under which they may be realised. The
final field configuration depends on the efficacy of both field-line
advection and the subsequent magnetic reconnection. These four scenarios are
summarised in figure \ref{flux_scenarios}.

In the first scenario for the phase of field rearrangement, fluid motions are effective at transporting the
magnetic flux in the superconducting shell to a small equatorial volume of the
shell. In the second scenario there is also large-scale fluid motion,
but this time it advects field lines towards the 
North and South poles. Finally, the last scenario (scenario 3)
accounts for the possibility of negligible fluid motion, and so the
field lines remaining undistorted as they cross the superconducting region.

After the advective phase, any reconnection then occurs. In
scenario 1, the flux has been concentrated
around the equator, forming an X-point geometry with neighbouring
field lines having the opposite sense from one another. If
reconnection is effective, such a geometry is able to reconnect fully,
with every field line pinching off across the incipient
superconducting shell, separating from the part of the field line on
the other side and rejoining a line on its own side (scenario 1i). If reconnection
is ineffective, the concentration of flux produced in an
equatorial band (scenario 1ii) will remain there, threading an
otherwise $B=0$ shell. But if the flux
is concentrated around the poles, reconnection cannot be effective,
both because neighbouring field lines will not have the opposite sense
needed, and because the characteristic lengthscale of the field across
the shell $\mathcal{L}_\perp$ is
too long to experience the requisite dissipation to allow for reconnection. The field must then continue to thread
through the superconducting shell in these two polar holes 
(scenario 2). In scenario 3, like scenario 2, there are no regions
with sufficiently distorted field lines to be subject to
reconnection; the field lines towards the symmetry axis are already
fairly straight, and those around the equator will tend to be `combed
out' by the boundary condition between the normal inner region and the
expanding superconducting shell (see \citet{hen_wass} for a detailed
discussion of this). All these possibilities are
plotted in figure \ref{flux_scenarios}.

Scenario 1i is what is usually understood by the term `Meissner effect', and
is what is implied in magnetic-field evolutions that implement a $B=0$
inner boundary and call it a `Meissner boundary condition'
(e.g. \citet{hollrued02,ponsgepp07}). It is
sometimes thought to be a consequence of type-I superconductivity, and 
this is not correct; a type-II superconductor can just as effectively
expel flux. However, it clearly requires a particular set of
circumstances to arise, even for $B<H_c$ where it represents the
minimum-energy configuration, because the physics may not allow for it
to be realised.
Scenario 3 is the opposite limit: one in which the macroscopic
magnetic field is essentially unaffected in both its magnitude and direction by
the developing superconducting region, except that the closed-field
line region tends to be pushed out of the core altogether, and into
the crust. In our model, all pre-condensation
fields $B_0\lesssim 10^{12}$ G or $B_0\gtrsim 5\times 10^{14}$ G are expected
to follow this evolutionary path and end up threading the entire
superconducting core, but even with the `right' initial field,
$10^{12}\,\mathrm{G}\lesssim B_0\lesssim 5\times 10^{14}\,\mathrm{G}$,
scenario 3 is also the likely outcome if there is limited advection of
field lines. It is
worth noting that equilibrium models for a type-II superconducting
core with $B<H_c$ \citep{roberts81,hen_wass,L14} also resemble this outcome, with no field lines
closing within the core itself.

Scenarios 1i and 3 are two extremes: one where field is entirely
expelled from the superconducting region, and another where it threads
the entire region. Between these two lie a variety of possibilities
for flux penetration in distinct, macroscopic regions, with other
regions where the field drops to zero; the two limiting cases of this
group of configurations are scenarios 1ii and 2.

We have argued that all of the four scenarios are, at the very least,
plausible, in that they place requirements on the lengthscale and
timescale of dynamics in the incipient shell which can be satisfied
for some part of the relevant parameter space. A detailed analysis
of the early evolutionary phase proposed here is well beyond
the scope of this paper, presenting a mixture of conceptual and
computational challenges (e.g. in modelling dynamics and reconnection
at the onset of superconductivity). Nonetheless, several aspects of
the different scenarios are amenable to calculation, using some
generic restrictions from MHD theory together with the notion of a
maximum permissible field strength, the critical field. We discuss
these next.

\subsection{Quantifying scenario geometries}

We are able to quantify, to some extent, the field borne from each of
our four scenarios, using the rules discussed in section \ref{restrictions}. We are
now focussed on the field just after any field rearrangement and
reconnection has taken place, whilst the shell is still thin; this
corresponds to the bottom row of configurations in the cartoon of
figure \ref{flux_scenarios}, though the thickness of the
superconducting shell there is increased for clarity and should not be
taken literally. In general both the magnetic field component normal
to the superconducting shell $B_\perp$ and the component tangential to
the shell $B_\parallel$ will vary with position, in different ways. As
sketched in the top panel of figure \ref{flux_scenarios}, the
pre-condensation field appears dominantly radial, i.e. $B_\perp\gg
B_\parallel$, but through advection (middle panel) it may generate a
substantial parallel component, resulting in $B_\perp\sim B_\parallel$.
Quantifying these effects requires a far more sophisticated analysis
than the scope of the present work. Instead, for simplicity will will
assume that $B_\perp$ is constant across the region of the shell that
it penetrates, and that the parallel component in these regions is a
fixed fraction of the normal component, i.e.
\beq
B_\parallel=\zeta B_\perp\ ,\ \ \zeta=\mathrm{const}.
\eeq

Of the four endpoints, we will start with scenario 2, where the
field cannot be fully expelled and reconnection will be ineffective
due to the large-lengthscale field geometry threading the polar
regions of the superconducting shell, with all neighbouring field
lines having the same sense. The simplest scenario is
that superconductivity minimises the volume of the $B\neq 0$ region of
the shell by squeezing the field lines inwards towards the
$\theta=0,\pi$ axis until the magnetic field is so concentrated that
its magnetic energy density reaches that of the
critical field, i.e. we are left with two normal-matter holes (one for
each pole) in the shell, where
\beq
\frac{B_\perp^2+B_\parallel^2}{8\pi}
=\frac{(1+\zeta^2)B_\perp^2}{8\pi}
=\frac{H_c^2}{8\pi}
\implies B_\perp=\frac{H_c}{\sqrt{1+\zeta^2}}.
\eeq
Given that we do not expect any significant reconnection, let us
assume that the magnetic
flux through both the inner and outer boundaries of the superconducting shell
$\mathfrak{F}_{\rm in}, \mathfrak{F}_{\rm out}$
is equal to the pre-condensation value $\mathfrak{F}_0$; see
equation \eqref{BC_rule_2}. Just after the onset of superconductivity
we may approximate the initial radius of onset of superconductivity
$\mathcal{R}$ and the inner and outer radii of 
the superconducting shell $R_{\rm in},R_{\rm out}$ as all having the same value,
and the conservation of flux expression becomes a relation between
the normal magnetic field component before $B_\perp^0$ and after
$B_\perp^{\rm post}$ condensation:
\beq
\mathfrak{F}_0
= 4\pi \mathcal{R}^2 B_\perp^0
= 2\pi \mathcal{R}^2 B_\perp^{\rm post}
\ 2\!\int\limits_0^{\theta_{\rm open}}\sin\theta\ \rmd\theta
=\mathfrak{F}_{\rm in}=\mathfrak{F}_{\rm out}\equiv\mathfrak{F}_{\rm post}
\eeq
Combining this result with the preceding equation to eliminate
$B_\perp^{\rm post}$ yields an expression for the
opening angle $\theta_{\rm pole}$ of the normal `hole' in the
superconducting shell:
\beq\label{theta_open}
\theta_{\rm open}=\arccos\brac{1-\frac{\sqrt{1+\zeta^2} B_\perp^0}{H_c}}.
\eeq
For this scenario, then, $\theta_{\rm pole}=\pi/2$ when the
pre-condensation field components are
$B_\perp=B_\parallel\approx 3.5\times 10^{14}\,\mathrm{G}$ -- i.e. flux
conservation dictates that any
field stronger than this must thread the entire initial shell of superconductivity.

Scenario 1ii is similar to that of scenario 2, except that now the
field is confined around the equator. Again, superconductivity acts to
minimise the volume of the shell where $B\neq 0$, resulting in a
single equatorial ring of normal matter, cleaving the superconducting
shell into two halves. The only change to the previous calculation is
that the limits on the $\theta$-integral must now be changed from
$\{0,\theta_{\rm open}\}$ to $\{\theta_{\rm eq},\pi/2\}$, where
$\pi/2-\theta_{\rm eq}$ is
the half-angular thickness of the ring (i.e. the total angular extent
is double this, by equatorial symmetry), and the result is:
\beq\label{theta_eq}
\theta_{\rm eq}=\arccos\brac{\frac{\sqrt{1+\zeta^2} B_\perp^0}{H_c}}.
\eeq
As expected, the same value of magnetic field as for scenario 2,
$B_\perp=B_\parallel\approx 3.5\times 10^{14}\,\mathrm{G}$, results in magnetic flux threading the
entire shell.

Scenario 1i invokes a reconnection event to `pinch off' all the
equatorial field lines and leave an unbroken $B=0$ superconducting
shell. Clearly flux is not conserved during this process; beforehand a
general field crossing the radius of onset of superconductivity would
be expected to have $B_\perp\sim B_\parallel$ and some flux
$\mathfrak{F}_0$, but afterwards $\mathfrak{F}_{\rm in}=\mathfrak{F}_{\rm out}=0$
and $B_\perp\to 0$ when approaching the superconducting shell from
either side, as steps in this component
violate $\div\bB=0$. $B_\parallel$, on the other hand, is
permitted to drop abruptly from its value in a normal domain to zero
in the superconducting shell, since steps in this component can be
matched with a surface current along the normal-superconducting
boundary. Flux conservation and the traction condition on $B_\perp$ do
not, therefore, provide any restriction on the geometry of scenario 1i.

Finally, in scenario 3 there is no motion tangential to the shell of
onset of superconductivity and so no distortion of field lines away
from their pre-condensation state. Once a thin superconducting shell has
formed, flux conservation and the traction condition imply that the
field lines must thread the entirety of the superconducting region on
a macroscopic scale, as
they previously threaded that volume when it was normal matter. As the
shell thickens, the field lines are thus combed out across the shell:
not strictly radial, which would lead to kinks at the inner and outer
shell boundaries and so require the existence of surface currents, but
with minimal curvature to ensure the whole length of each field line
is smooth. On a microscopic scale, however, the flux distribution will
differ from the pre-condensation state. The evenly-distributed flux in
the normal-matter region will become split up into thinner flux
structures in the superconducting shell, over a transition region
whose thickness is the penetration depth. Depending on the type of
superconductivity operative in the shell, the flux structures will
either be a predictable Abrikosov lattice of evenly-distributed
fluxtubes with central magnetic field 
strength $B=H_{c1}$ surrounded by $B=0$ matter (type-II
superconductivity), or regions with various possible thicker (but still
small-scale) flux structures with $B=H_c$ alternating with $B=0$
regions. These two states are the favoured, minimum-energy, states for
flux penetration when $H_{cI}<B<H_c$ (type-I superconductivity) or
$H_{c1}<B<H_{c2}$ (type-II superconductivity), whereas for the weaker
fields we consider here, the global minimum energy state would be a
complete shell of $B=0$. For this region, our scenario 3 represents a
state that is sometimes dubbed `quasi-stable', but we would argue that
there is no need for the qualifying prefix `quasi': it is simply
stable, being a local minimum-energy state from which there is no
feasible route for the system to reach the $B=0$ global minimum.
We believe this to be a more solid argument than earlier ones
(e.g. \citet{BPP69a}) for why
a neutron star with a relatively weak field has flux penetrating the entire core rather than
exhibiting the Meissner effect.

\subsection{Energy release during reconnection}
\label{E_recon}

Let us denote the pre-condensation magnetic energy as
$\mathcal{E}^0_{\mathrm{mag}}$, and use
$\mathcal{E}^{\mathrm{str}}_{\mathrm{mag}}$
to denote the additional magnetic energy gained from
the stretching of field lines in the initial shell of
superconductivity. Since in practice we will never be able to
`measure' the pre- or post-condensation magnetic energies, the more
relevant quantity is the fractional change $\alpha$ in magnetic energy
resulting from the advection of field lines at the onset of condensation, i.e.
\beq
\alpha\equiv\frac{\mathcal{E}^{\mathrm{str}}_{\mathrm{mag}}}{\mathcal{E}^0_{\mathrm{mag}}}.
\eeq
Reconnection takes a magnetic field from a higher-energy configuration
to one of lower energy, so this process always releases energy; we
will assume this is some unknown fraction $\beta$ of the energy gained
during the advection phase
$\mathcal{E}^{\mathrm{str}}_{\mathrm{mag}}$. How much depends on
various factors such as how 
extensive field-line stretching is, and the initial
field geometry, but for any large-scale field rearrangement the
fraction will be significant, so we suggest $0.1<\beta\lesssim
1$. Note that $\beta>1$ is also quite possible: it means the
Meissner-expelled state is of lower energy than the pre-condensation
field $\bB_0$. Recall that
reconnection will not occur for a pre-condensation field
less than $\sim 10^{12}\,\mathrm{G}$. In addition, the requisite
field-line-stretching needed to produce sharp features on which
reconnection can act will not occur if the pre-condensation field
$\gtrsim H_c$.
The value of $B_0$ cannot be converted to a precise pre-condensation magnetic energy
because this depends on the quantitative field structure, but we can
approximate it as some average magnetic energy density $\bar{B}_0^2/8\pi$
multiplied by the volume of the star,
\beq
\mathcal{E}^0_{\mathrm{mag}}
\approx\frac{\bar{B}_0^2R_*^3}{6}
= 2.9\times 10^{41}\brac{\frac{\bar{B}_0}{10^{12}\,\mathrm{G}}}^2
                 \brac{\frac{R_*}{12\,\mathrm{km}}}^3\ \mathrm{erg}.
\eeq
Summarising the arguments of this subsection, the kind
of reconnection event required to produce a Meissner-expelled magnetic
field will release somewhere in the range
\beq\label{E_release}
\mathcal{E}_{\mathrm{mag}}\approx
\alpha\beta\,\times (2.9\times 10^{41}-7.2\times 10^{46})\,\mathrm{erg}
\eeq
of energy, where the parameters $\alpha,\beta$ are most likely to be of order
unity. In the following section we will make quantitative calculations
of the former. The above energy release will coincide with the
onset of superconductivity, i.e. after a few minutes. Although the
supernova is still very bright at this stage, this rather specific
signature could potentially be detectable -- or at least
constrained. Furthermore, exactly when it occurs would provide a 
valuable constraint on the variety of different energy gap models, as
it would lead directly to an estimate of the maximum value of $T_c$
\citep{meissner_letter}. Although enhanced cooling is also a signature
of the start of Cooper pairing, and neutron-neutron pairing may be
responsible for the current thermal
evolution of the Cas A neutron star (whose age is around
$350\,\mathrm{yr}$) \citep{page11,ho15} , the
corresponding effect due to proton-proton pairing would occur so close
to birth as to be masked by the still-bright supernova. The possible
energy injection discussed here is therefore the only plausible
possibility for observationally constraining the proton gap model.

\section{A quantitative model for field rearrangement}

Any of the above scenarios could, potentially, be realised, with the
details depending on the pre-condensation magnetic field, the internal
fluid motion of the star, and the efficiency of reconnection
mechanisms. We now consider a simple concrete example where all
calculations are semi-analytic, to get some quantitative results.

The most general axisymmetric poloidal magnetic field
satisfying the solenoidal constraint $\div\bB=0$ takes the form
\beq\label{u_defn}
\bB=\nabla u\times\be_\phi,
\eeq
where $u=u(r,\theta)$ is the poloidal streamfunction.
Note that from equation
\eqref{u_defn},
\beq
\bB\cdot\nabla u=0,
\eeq
which implies that $u=\mathrm{const}$ along a given field line. Every field line can therefore be labelled
the value $u=u_0$ along it, which will be convenient later. This value
lies in the range $0\leq u\leq u_{\rm max}$, where $u=0$ for
$\theta=0$ and
$u_{\rm max}$ is attained at the centre of the equatorial region of closed
field lines, at which the poloidal magnetic field strength drops to
zero.

\begin{figure*}
\begin{center}
\begin{minipage}[c]{0.4\linewidth}
\includegraphics[width=\linewidth]{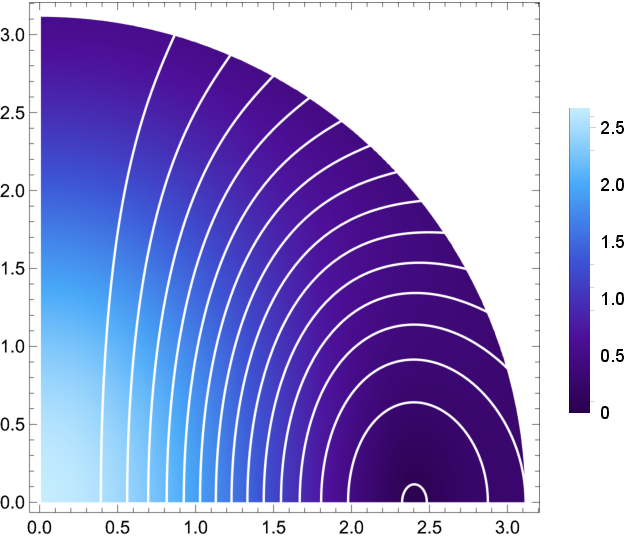}
\end{minipage}
\caption{\label{monaghan}
 A purely poloidal field in hydromagnetic equilibrium in an $N=1$
 polytropic star. The surface of the star
 is at a radius $r=\pi$, in the usual dimensionless units used in the
 Lane-Emden equation. Colourscale shows the magnitude of the field;
 white lines show contours of constant value of the streamfunction,
 equivalently, the magnetic field lines.}
\end{center}
\end{figure*}

In the case of a
dipolar magnetic field in magnetohydrostatic equilibrium in an $N=1$
polytropic star, \citet{monaghan65} found the following analytic solution
for the streamfunction:
\beq\label{monaghan_u}
u=\mathfrak{B}\sin^2\theta
\brac{\frac{4\sin\xi}{\xi}-4\cos\xi-2\xi\sin\xi-\frac{2\xi^2}{3}}
\eeq
where $\mathfrak{B}$ is an arbitrary constant setting the magnitude of
the magnetic field and $\xi$ is the usual dimensionless radius from
the solution to the Lane-Emden equation \eqref{sinc}. We plot the
magnitude and direction of the resulting magnetic field in figure
\ref{monaghan}. Note that the inclusion of more realistic equations of state and thermal
pressure contributions result in only minor deviations from this
simple model \citep{hoteqm}.

To evaluate the magnetic energy along a fluxtube, equation
\eqref{emag_ft_orig}, we need a way to parametrise position along a given
field line. We cannot use the streamfunction $u$, because a field line
is defined by $u=\mathrm{const}$. Each field line is a set of 2D
coordinates $\{r,\theta\}$, so if we can express either of the
coordinates as a function of the other, i.e. $r=r(\theta)$ or
$\theta=\theta(r)$, this will give us a satisfactory parametrisation
to use.

Looking at figure \ref{monaghan} and
imagining drawing spokes radially outwards from the origin (i.e. lines of
$\theta=\textrm{const}$), we see
that in general (it is easiest to visualise in the equatorial
region of closed field lines), a single spoke
passes through the same field line in two places, so writing
$r=r(\theta)$ would map one radial value to \emph{two} positions on
the field line, and therefore $r(\theta)$ violates the definition of a
function and we cannot use $\theta$ for location along a given field
line $u=u_0$. Instead imagining concentric circles of constant $r$, we see
that each of these crosses a single field line in just one location;
reversing the previous logic means that we \emph{can} parametrise position
along any given field line $u=\mathrm{const}\equiv u_0$ by $r$. We
may now evaluate the magnetic energy along a fluxtube, equation
\eqref{emag_ft_orig}, but it is informative to start with the simpler
calculation of the length of a field line, stretched and
unstretched. We present the details for the case of advecting field
lines towards the equator, as in figure \ref{stretched_field_line};
the case of stretching field lines towards the pole is a trivial
modification to the calculation (and we will give results for this
case too).

Using a standard result for lengths of parametrised curves, an
unstretched field line $u=u_0$ has length
\beq\label{L1_eqn}
L_1(u_0)=2\int\limits_{r_{\rm min}}^{r_{\rm max}}\sqrt{1+r^2
  \brac{\td{\theta(r,u_0)}{r}}^2}\ \rmd r,
\eeq
where the integration is carried out only in the quadrant shown in
figure \ref{monaghan}, exploiting the equatorial symmetry of the
field, so the factor of $2$ gives the full length of
the field line. The length of the additional stretched component is
simply twice that of the circular arc from where the field line intersects the
radius $r=\mathcal{R}$ to the equator $\theta=\pi/2$, i.e.
\beq\label{L2_eqn}
L_2(u_0)=2\mathcal{R}\brac{\frac{\pi}{2}-\theta(\mathcal{R},u_0)}.
\eeq

\begin{figure}
\begin{center}
\begin{minipage}[c]{0.5\linewidth}
\includegraphics[width=\linewidth]{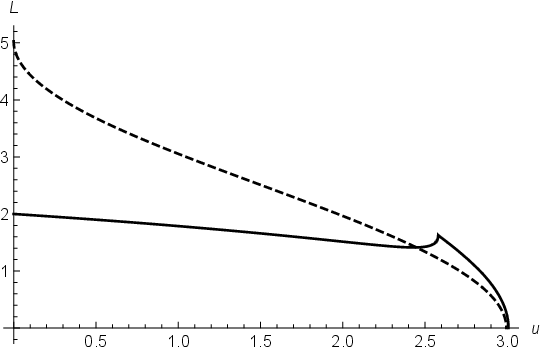}
\end{minipage}
\caption{\label{L1_L2}
Length of unstretched field lines $L_1$ (solid line) and the additional
stretched component $L_2$ (dashed line), in dimensionless
units divided by $\pi$, as a function of the constant value of the
streamfunction labelling each field line, from the straight field line
running from pole to pole ($u_0=0$), past the start of the closed
field line region $u_0\approx 2.6$ to the maximum value at the centre
of the closed-field-line region $u_0\approx 3.0$.}
\end{center}
\end{figure}

We plot $L_1,L_2$ for field lines throughout the star in figure
\ref{L1_L2}. To understand this plot, recall that every field line can be labelled
the value $u=u_0$ along it. The field line with $u=0$ is the straight
line along the magnetic-field symmetry axis, which runs from pole to
pole. Moving away from this axis $u$ increases\footnote{Actually, as
  defined in the Monaghan solution,
  $u$ drops from zero along the axis to being negative definite
  everywhere else. When referring to specific values of the
  streamfunction we actually quote the absolute value $|u|$
  to avoid confusion when referring to the `maximum' value $u_\mathrm{max}$.}, reaching the value
$u_0=(2\pi^2-12)/3$ at which field lines begin to close inside the
star. Moving further into the closed-field-line region $u$ continues
to increase, and reaches a 
maximum $u=u_{\rm max}\approx 3.0069$ at the centre of the closed-field line
region. In the dimensionless units used, the radius of the star is at
a value $\pi$. We therefore expect the straight field line running
from pole to pole to have a length equal to the diameter of the star,
$2\pi$, and for the length to drop to zero at the centre of the
closed-field-line region; this is seen in figure \ref{L1_L2} for the
unstretched field line $L_1$. There is not a monotonic decrease in
field-line length from $u=0$ to $u=u_{\rm max}$: a small cusp is seen
corresponding to the transition from open to closed field lines (only
the length inside the star is measured, otherwise it would be a smooth
decrease at the transition). If the field lines are maximally
stretched towards the equator and back at the radius of onset of
superconductivity fixed at $\mathcal{R}=0.8R_*$, as shown in figure
\ref{stretched_field_line}, the additional length is
the value $L_2$ plotted. The field line running along the symmetry
axis should have its length increased by $0.8\pi\times 2\pi\approx
5.0\pi$, which agrees with the plot, and this stretched component
decreases to zero for field lines progressively closer to the equator
(which therefore have less distance to cover), as expected.

\begin{figure}
\begin{center}
\begin{minipage}[c]{\linewidth}
\includegraphics[width=0.5\linewidth]{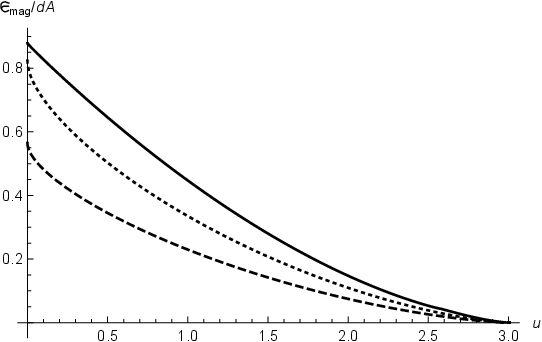}
\includegraphics[width=0.5\linewidth]{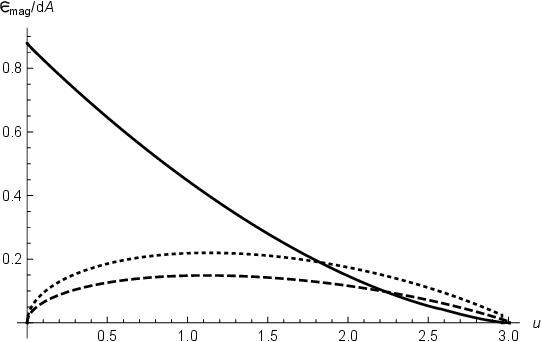}
\end{minipage}
\caption{\label{B_incr}
Energy per field line as a function of value $u=u_0$. Unstretched
field shown with the solid line, dashed (dotted) lines show the
additional magnetic energy for each field line due to the stretched
piece of the field when $\mathcal{R}=0.8\pi$ ($0.7\pi$). Left-hand
panel shows the increase in energy by stretching field lines towards
$\theta=\pi/2$; right-hand panel shows the corresponding energy
increases resulting from stretching field lines towards $\theta=0$.}
\end{center}
\end{figure}

Stretching magnetic field lines increases the energy of the
configuration; recall section \ref{field_stretch}. It is informative to think of this process through the
contributions of individual field lines, or more specifically a thin
volume with length given by the magnetic-field line and some
infinitesimal cross-sectional area $dA$. We will refer to the integral
of the magnetic energy density $B^2/8\pi$ along a field line
multiplied by $dA$ as the
`magnetic energy of the field line' and denote this
$\mathfrak{E}_{\rm mag}$.
For the Monaghan field, combining the contributions from the
unstretched piece $\mathfrak{E}_{\rm mag}^0$ and the stretched region
$\mathfrak{E}_{\rm mag}^{\rm str}$ for a given field line
$u=u_0$ gives us:
\beq
\mathfrak{E}_{\rm mag}
=\mathfrak{E}_{\rm mag}^0+\mathfrak{E}_{\rm mag}^{\rm str}
= \left\{2\int\limits_{r_{\rm min}}^{r_{\rm max}}[\frac{B(r,\theta(r))]^2}{8\pi}
                                                    \sqrt{1+r^2  \brac{\td{\theta(r,u_0)}{r}}^2}\ \rmd r
     \ \ +\ \ 4\mathcal{R}\brac{\frac{\pi}{2}-\theta(\mathcal{R},u_0)
                    \frac{[B(\mathcal{R},\theta(\mathcal{R},u_0))]^2}{8\pi}}
                \right\} \times dA
\eeq
where we have combined equations
\eqref{emag_ft_stretch}, \eqref{L1_eqn} and \eqref{L2_eqn}. In the
calculations that follow we set the prefactor $\mathfrak{B}$ from
equation \eqref{monaghan_u} to unity. This results in no loss of
generality, as we are interested in the fractional increase in
magnetic energy per field line for a given starting magnetic field, so the $\mathfrak{B}$
factors cancel from numerator and denominator in such a ratio. There
is also no qualitative difference between the field-line structure at
different field strengths.

Figure \ref{B_incr} shows that the increase in $\mathfrak{E}_{\rm mag}$ is not
simply proportional to the increased length of the field line, because
the field strength at the point where the superconducting shell begins
varies across the star. For $\mathcal{R}$ further in, the total
increase in field line length is reduced, but because the field
strength is higher towards the centre, the magnetic energy per field
line is actually
increased. We notice that the energy increases are far greater if
field lines are stretched towards the equator rather than the
pole. This is because in the former case the field lines that are most
stretched are those in regions of strong magnetic field; in the latter
case the most stretching is from the closed-field-line region, where
$B$ is small (and drops to zero in the centre).

\begin{table*}
\begin{center}
\begin{tabular}{ccccccc}
    position of $\mathcal{R}$  & $0.9$ & $0.85$ & $0.8$ & $0.79$ & $0.75$ & $0.7$ \\
    \hline
  $\mathcal{E}_{\rm mag}$ increase from field lines stretched towards equator
       & $0.5380$ & $0.6267$ & $0.7212$ & $0.7411$ & $0.8233$ & $0.9271$\\
  $\mathcal{E}_{\rm mag}$ increase from field lines stretched towards pole
       & $0.5029$ & $0.5436$ & $0.5930$ & $0.6058$ & $0.6704$ & $0.7817$\\
\end{tabular}
\caption{\label{Emag_incr}
  Fractional increase in magnetic energy,
  $\mathcal{E}_{\rm mag}^{\rm str}/\mathcal{E}_{\rm mag}^0$, as a function of
position of superconducting shell $\mathcal{R}$. We consider a
plausible range of values of $\mathcal{R}$ in increments of $0.05R_*$,
and also the `fiducial' model with $\mathcal{R}=0.79R_*$. In the
dimensionless units used in the calculation, the pre-condensation
magnetic energy $\mathcal{E}_{\rm mag}^0=5.4937$.}
\end{center}
\end{table*}

This calculation is in dimensionless units, but because the
magnetic-field structure is the same at any field strength, the
results can be readily rescaled. The key result will be quantifying the fractional
increase in total magnetic energy $\alpha$ in the two cases where the field is
stretched in the onset shell of superconductivity. Note, however, this
quantitative calculation adds nothing new to our understanding of the
efficacy of reconnection i.e. the $\beta$ parameter from equation
\eqref{E_release} describing the amount of magnetic energy that is
ultimately released.

We now want to understand the total increase in magnetic energy when all
the field lines across the shell $r=\mathcal{R}$ are suitably
distorted towards either the equator or pole. Similarly to the calculation for the length of a field
line, equation \eqref{L1_eqn}, we may also calculate the area of the surface
of revolution obtained from rotating this field line through an angle
of $2\pi$ radians in the azimuthal direction:
\begin{align}
S_1(u_0)&=2\int\limits_{r_{\rm min}}^{r_{\rm max}}
                 2\pi x(r)\sqrt{1+r^2\brac{\td{\theta(r,u_0)}{r}}^2}\ \rmd r\nn\\
             &=4\pi\int\limits_{r_{\rm min}}^{r_{\rm max}}
                 r\sin\theta(r)\sqrt{1+r^2\brac{\td{\theta(r,u_0)}{r}}^2}\ \rmd r.
\label{S1_eqn}
\end{align}
Finally, the whole volume of the star can then be described as a set
of nested surfaces of revolution, each with constant $u$ and with an
infinitesimal spacing $du$ between them -- that is, the volume of the
star $V$ is given by the integral
\beq\label{V_uint}
V=\int\limits_{u=0}^{u_{\rm max}}S_1(u)\ du.
\eeq
Similarly, we will find the magnetic energy of the original
magnetic field $\mathcal{E}_{\rm mag}^0$ and the additional stretched
component $\mathcal{E}_{\rm mag}^{\rm str}$ by first evaluating the
following functions (note that these differ from the
$\mathfrak{E}_\mathrm{mag}$ functions in their extra
$2\pi r\sin\theta(r)$ surface area element in the integrand) for a given field line $u=u_0$:
\begin{align}
E_0(u_0)&=\frac{1}{2}\int\limits_{r_{\rm min}}^{r_{\rm max}}
           B^2 r\sin\theta(r)\sqrt{1+r^2\brac{\td{\theta(r,u_0)}{r}}^2}\ dr,\\
  E_{\rm str}(u_0)&=8\pi\mathcal{R}^2\sin\theta(\mathcal{R},u_0)
                    \brac{\frac{\pi}{2}-\theta(\mathcal{R},u_0)
          \frac{[B(\mathcal{R},\theta(\mathcal{R},u_0))]^2}{8\pi}},\\
\end{align}
and then integrating these over $u$:
\begin{align}\label{Emag0_uint}
\mathcal{E}_{\rm mag}^0
   & = \int\limits_{u=0}^{u_{\rm max}}E_0(u)\ du,\\
\mathcal{E}_{\rm mag}^{\rm str}
  & = \int\limits_{u=0}^{u_{\rm max}}E_{\rm str}(u)\ du,\\
\label{Emagstr_uint}
\end{align}

In practice, we find consistent inaccuracies with this approach. The
volume calculated from equation \eqref{V_uint} is $4.8\%$ greater
than the exact value of $4\pi^4/3$ for a sphere of radius
$R_*=\pi$, and the magnetic energy for the pre-condensation field calculated
in the same manner, from equation \eqref{Emag0_uint}, is $29\%$
greater than the straightforward volume
integral\footnote{In fact the magnetic energy evaluated both in this conventional way and via the
field-line parametrisation should both take the radial integration out
to infinity; but the result of stopping at the surface should be the
same for both methods.}
of the energy density:
\beq
\frac{1}{8\pi}
\int\limits_{\phi=0}^{2\pi}
\int\limits_{\theta=0}^{\pi}
\int\limits_{r=0}^{R_*}
       B^2r^2\sin\theta\ \rmd r\rmd\theta\rmd\phi.
\eeq
These discrepancies seem to stem from a shortcoming of
parametrising a field line as $r=r(\theta)$, since although this is
formally a satisfactory choice (as discussed earlier), towards the
equator field lines become tangential to contours of constant radius
-- so a substantial change in field-line location becomes a very small
change in radius. To check whether the parametrisation was at fault or
an error in our calculations, we tried a different magnetic field 
configuration with straight vertical lines, and in this case the
integral of field-line contributions over 
the range of values of the streamfunction \emph{did} yield the
correct results for volume and magnetic energy. This gives us
confidence that the method of calculation is not itself at fault and,
although the resulting $29\%$ inaccuracy in the energy is undesirable,
it is likely to affect the calculations of both $\mathcal{E}_{\rm mag}^0$
and $\mathcal{E}_{\rm mag}^{\rm str}$ in a similar manner. Since we
are interested in the fractional increase in magnetic energy,
$\alpha=\mathcal{E}_{\rm mag}^{\rm str}/\mathcal{E}_{\rm mag}^0$, the
systematic inaccuracies in both numerator and denominator are likely
to cancel each other out to a large extent, resulting in a ratio that
is reliable to our order of working.

Returning to the arguments of section \ref{E_recon}, any reconnection event
will be accompanied by the release of a substantial amount of
energy. To quantify this we now need to convert from dimensionless to
cgs units. Using the fact that the
dimensionless radius $\hat{r}$ and magnetic field at the polar cap
$\hat{B}_{\rm p}$ are given by
\beq
\hat{r}=\frac{\pi r}{R_*}\ ,\
\hat{B}_{\rm p}=\frac{B_{\rm p}}{\mathfrak{B}} =\frac{4}{3}-\frac{8}{\pi^2},
\eeq
we simply need to multiply these quantities by a prefactor to yield the
physical ones:
\beq
r=3.2\times 10^5 \hat{r}R_{12}\ \mathrm{cm}\ ,\
B_{\rm p}=1.9\times 10^{12} \hat{B}_{\rm p} B_{12}\ \mathrm{G}.
\eeq
Two other useful relations are obtained from the dimensionless
magnetic energy $\hat{\mathcal{E}}^0_\mathrm{mag}=5.4937$. We can use
it to define an average internal magnetic field $\hat{\bar{B}}$, and
hence find the ratio of polar-cap to average internal field, which is
the same in both dimensionless and physical units:
\beq
\frac{\bar{B}}{B_\mathrm{p}}=\frac{\hat{\bar{B}}}{\hat{B}_\mathrm{p}}
 = \frac{1}{\hat{B}_\mathrm{p}}\sqrt{\frac{\hat{8\pi\mathcal{E}}^0_\mathrm{mag}}{V}}
 = 1.972.
\eeq
We can also calculate magnetic energy from equation \eqref{Emag0_uint} in physical units, which entails
multiplying by $B_{\rm p}^2 R_*^3$, giving us:
\beq
\mathcal{E}^0_{\mathrm{mag}}
= 6.5\times 10^{41}\brac{\frac{B_\mathrm{p}^0}{10^{12}\,\mathrm{G}}}^2
                 \brac{\frac{R_*}{12\,\mathrm{km}}}^3\ \mathrm{erg}.
\eeq
For simplicity let us assume that all the additional energy
from field-line stretching is released, i.e. $\beta=1$. 
In the case of the stretch towards the equator, full reconnection is
possible, and depending on where the initial shell of
superconductivity $\mathcal{R}$ is located (which in turn depends upon the gap
model), the value of $\alpha$ will differ; (see table
\ref{Emag_incr}). This leads to a predicted energy release in the range
\beq
\mathcal{E}^{\mathrm{out}}_{\mathrm{1i}}
=(3.5-6.0)\times 10^{43}\, B_{\mathrm{p},13}^2\,\mathrm{erg}.
\eeq
For our fiducial model with $\mathcal{R}=0.79$, the prefactor in the
above expression is $4.8$. In the case of stretching towards the
pole, however, we do not expect substantial reconnection -- let us
take $\beta=0.1$ for the sake of definiteness -- and table \ref{Emag_incr} shows
us that $\alpha$ is also a little smaller in this case, and so we estimate
\beq
\mathcal{E}^{\mathrm{out}}_2
=(0.33-0.51)\times 10^{43}\, B_{\mathrm{p},13}^2\,\mathrm{erg}
\eeq
in this latter case, with a prefactor of $0.39$ for the fiducial
model. However such an energy release manifests itself, it will
clearly produce a rather weaker signal. Finally, scenarios 1ii and 3 both
assume reconnection is at best feeble, and so in these cases negligible energy release would be
expected at the onset of superconductivity.

\begin{figure}
\begin{center}
\begin{minipage}[c]{0.4\linewidth}
\includegraphics[width=\linewidth]{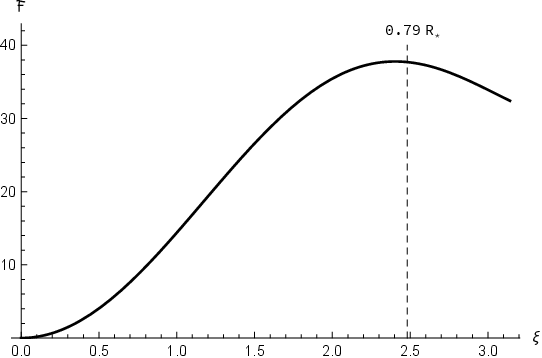}
\end{minipage}
\caption{\label{flux_vs_radius}
Magnetic flux through a spherical surface of constant radius, as a
function of dimensionless radius $\xi$ in the star. The onset of superconductivity in
our fiducial model is marked.}
\end{center}
\end{figure}

Earlier on we used the requirement of flux conservation together with
$B\leq H_c$ to calculate
the angular extent of any $B\neq 0$ hole through the otherwise
Meissner-expelled superconducting shell, culminating in equations
\eqref{theta_open} and \eqref{theta_eq}. This general treatment of flux conservation can now also be made
specific to the Monaghan field. For the post-rearrangement phase we made the assumption that
$B_\perp^2=B_\parallel^2$, but one can see visually that for the
specific model considered here, the field is virtually radial at a
radius of $0.79\pi$ (and near to the equator, where the angular
component is relatively large, the overall magnitude is close to
zero). We quantify the typical value of $B_\parallel^2/B_\perp^2$ with the ratio
\beq
\frac{\int_0^{\pi/2}B_\parallel^2(\mathcal{R},\theta)\, d\theta}
        {\int_0^{\pi/2}B_\perp^2(\mathcal{R},\theta)\, d\theta}
=0.0048\textrm{ for }\mathcal{R}=0.79\pi.
\eeq
The criterion for overcoming superconductivity, written in terms of
$B_\perp$, then becomes
\beq
B_\perp=\frac{H_c}{\sqrt{1.0048}}
 = H_c \textrm{ to our order of accuracy.}
\eeq 
Otherwise we follow the same working as before, which for scenario
2 leads to a polar hole opening angle of
\beq\label{theta_open_Mon}
\theta_{\rm open}
=\arccos\brac{1-\frac{\mathfrak{F}_0}{4\pi\mathcal{R}^2 H_c}}
\eeq
and for scenario 1ii, an equatorial ring of half-angular thickness
\beq\label{theta_eq_Mon}
\theta_{\rm eq}
=\arccos\brac{\frac{\mathfrak{F}_0}{4\pi\mathcal{R}^2 H_c}}.
\eeq

\begin{figure}
\begin{center}
\begin{minipage}[c]{0.5\linewidth}
\includegraphics[width=\linewidth]{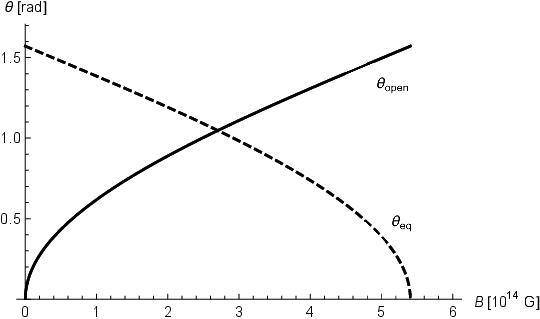}
\end{minipage}
\caption{\label{theta_open_eq}
Opening angles for $B\neq 0$ regions within the superconducting shell, in
scenarios 1ii ($\theta_{\rm eq}$, the dashed line)
and 2 ($\theta_{\rm open}$, the solid line), as a function of field 
strength. $\theta_{\rm open}$ increases with field strength until it
reaches the equator. $\theta_{\rm eq}$ begins at the equator,
$\theta=\pi/2$, so for increasing field strength its value decreases,
as it expands out towards the symmetry axis running from pole to pole.}
\end{center}
\end{figure}

Were the field strength in the inner and outer regions to remain
constant as the superconducting region expands, then by flux conservation through the polar hole (or
equatorial ring), the $B\neq 0$ zone within the superconducting shell would be of
constant cross-sectional area, i.e. a cylindrical hole in the case of
scenario 2. However, because the expansion of the superconducting
shell also decreases the volume of the inner and outer normal regions,
flux conservation also implies that the field will be `concentrated'
and amplified. We explore this next.

\section{Later evolution}
\label{later_evol}

\subsection{Field evolution in NS cores}

The expansion of the superconducting shell, in every case except
scenario 3, drives an amplification of the magnetic field in the
$B\neq 0$ regions, as described in the following subsection. The timescale for this process
depends on the specific superconducting gap profile, but at least the
early phase should proceed over a mere few years after birth, considerably faster than any other
mechanism for core-field evolution. At later times, if the critical
temperature drops significantly towards the centre of the star, the
Meissner-induced field amplification may slow down considerably, and
it becomes important to understand whether this process might work in
tandem with other field-evolution processes in the core.

Core field evolution is a contentious
topic, with a variety of approaches that lead to wildly different
estimates for the characteristic evolutionary timescale: as short as $10^3\,\mathrm{yr}$
(e.g. \citet{castillo20}) and at least as long as
$10^{11}\,\mathrm{yr}$ (e.g. \citet{graber15}). This is not our
primary focus here, but a review of recent literature suggests that
the only short-timescale mechanisms assume normal matter
(i.e. non-superconducting protons, non-superfluid neutrons)
\citep{castillo20,moraga24}. Earlier work suggesting short-timescale
evolution in a superconducting core (e.g. \citet{jones06}) has
recently been criticised, with counter-arguments suggesting far slower
evolution \citep{pass17,gusakov19}, at least if the star is static
\citep{gusakov20}.

We conclude that the dynamic Meissner effect considered in
this paper will, in general, not add to this discordance, as it should
be completed well before even the fastest plausible core-field
evolution mechanisms have begun their work. In particular, for the
specific approximate critical temperature profile we adopt,
the transition to superconductivity is completed (in the regions
where it \emph{can} be completed) in a matter of a few decades.

\subsection{Compression of flux}

Whichever of our four scenarios for field rearrangement (recall figure
\ref{flux_scenarios}) is realised, the later evolution will be
dictated by this initial shell formation. If flux penetrates the
shell, it will continue to do so as the shell expands, a process which
we will assume, for simplicity, does not change the field structure
itself within that region (a related issue is touched upon in section
\ref{tor_mix_fields}). In all cases except scenario 3, the volume of the star
threaded by the macroscopic field decreases as the superconducting
shell expands. In ideal MHD, however, Alfv\'en's frozen-flux theorem
dictates that expulsion of magnetic flux would have to be accompanied
by expulsion of the fluid too! Instead, Alfv\'en's theorem can only be
valid for the normal proton fluid, which is strongly coupled to the
(non-superfluid) neutron fluid too. Over a transition region into the
superconducting shell, where $T\lesssim T_c$ but
$T\ll\!\!\!\!\! / \ \, T_c$,
the proton fluid will behave like an admixture of normal and
superconducting components, as for the two-fluid model of liquid
helium; fluid elements will effectively start to be able to slip
across field lines rather than being permanently threaded by them.
This phase deserves further study; for now we simply point out that
without this re-definition of Alfv\'en's theorem, there can be no
development of the Meissner effect in the thickening superconducting shell.

With this revised Alfv\'en's theorem, the expansion of the
superconducting shell pushes field lines beyond the outer boundary $r=R_\mathrm{out}$
further outwards, and inner field lines $r<R_\mathrm{in}$ further inwards. We can see by
flux conservation that this will necessarily amplify the magnetic
field. More precisely, since the expansion of the superconducting
shell is radial, we will also assume the magnetic field is pushed
radially inwards/outwards. We cannot apply Eulerian flux conservation, i.e.
across a shell of fixed radius, because the region hosting the flux is
shrinking from its initial radial extent of $\mathcal{R}$ for the
inner sphere ($R_*-\mathcal{R}$ for the outer shell). Instead we need
to consider Lagrangian flux conservation. Let $R_1$ be the radius of
some arbitrary
spherical shell at the onset of superconductivity,
$R_{\rm in},R_{\rm out}$ the inner and outer radii of the
superconducting shell some time later, and $\mathcal{R}$ the radius of
the initial shell of superconducting matter. Then Lagrangian flux
conservation means that the magnetic flux through the shell $r=R_1$ at
the onset of superconductivity should be equal to the flux through a
\emph{different} shell $r=R_2$ once the superconducting shell has
expanded, where
\beq
\label{R1_to_R2}
R_2=
\begin{cases}
  \frac{R_{\rm in}}{\mathcal{R}} R_1\ &\textrm{ for }R_1<\mathcal{R},\\
  R_*-\frac{(R_*-R_{\rm out})}{(R_*-\mathcal{R})}(R_*-R_1)\ &\textrm{ for }R_1>\mathcal{R}.
\end{cases}
\eeq
Flux conservation is then expressed as
\begin{align}
\mathfrak{F}_1
  \equiv R_1^2\int B_r(R_1,\theta,\phi)\ d\theta d\phi
  =R_2^2\int B_r(R_2,\theta,\phi)\ d\theta d\phi
\equiv\mathfrak{F}_2
 \implies
  \frac{\int B_r(R_2,\theta,\phi)\ d\theta d\phi}{\int B_r(R_1,\theta,\phi)\ d\theta d\phi}
= \brac{\frac{R_1}{R_2}}^2
\ &\textrm{ for }R_1<\mathcal{R},\\
  \frac{\int B_r(R_*-R_2,\theta,\phi)\ d\theta d\phi}{\int B_r(R_*-R_1,\theta,\phi)\ d\theta d\phi}
= \brac{\frac{R_*-R_1}{R_*-R_2}}^2
\ &\textrm{ for }R_1>\mathcal{R}.
\end{align}
In the special case where the radial magnetic field is a separable
function, of the form $B(r,\theta,\phi)=f(r)g(\theta)h(\phi)$ -- which
includes the Monaghan field -- the above expressions reduce to
considerably simpler forms:
\beq\label{flux_compression}
\frac{B_r(R_2)}{B_r(R_1)} 
= \brac{\frac{R_1}{R_2}}^2
\ \textrm{ for }R_1<\mathcal{R}\ ,\ \ \ \ 
\frac{B_r(R_*-R_2)}{B_r(R_*-R_1)} 
= \brac{\frac{R_*-R_1}{R_*-R_2}}^2
\ \textrm{ for }R_1>\mathcal{R}.
\eeq
Note that we have to be careful to choose a shell \emph{within} one of the
normal-matter regions, because along the boundary shells at
$r=R_{\rm in},r=R_{\rm out}$ the flux is zero, so flux conservation
here does not give any information about the amplification of $B$ due
to the expansion of the superconducting shell. Finally, although we
have found a relation for the amplification of $B_r$, if we assume the
field geometry is unchanged by this compression, then the other
components of $\bB$ must also be magnified in the same way. On the
other hand, even if the other field components are assumed \emph{not} to be
amplified in the same way, they must still be excluded from the superconducting
shell -- ingredients for the possible (and perhaps inevitable)
development of current sheets; we will therefore dismiss this
possibility, and so regard the previous relation as applying to the
magnitude of the total magnetic field and not just its radial
component. We can then combine equations \eqref{R1_to_R2} and
\eqref{flux_compression} to give
\beq\label{amplify_B}
B(R_2)  = \brac{\frac{\mathcal{R}}{R_{\rm in}}}^2\ B(R_1)
\ \textrm{ for }R_1<\mathcal{R}\ ,\ \ \ \ 
B(R_*-R_2)  = \brac{\frac{R_*-\mathcal{R}}{R_*-R_{\rm out}}}^2\ B(R_*-R_1)
\ \textrm{ for }R_1>\mathcal{R}\ .
\eeq

\subsection{Final state after flux compression}

The flux compression described above cannot continue indefinitely,
as it will eventually result in a magnetic field with $B=H_c$. At this
point the expansion of the superconducting shell is arrested, and the
final configuration from scenarios 1i, 1ii and 2 is dictated by this. Along the way to reaching this final
state, as the magnetic field is amplified, this also affects the
opening angle of the polar holes/equatorial ring from scenarios 1ii
and 2. We discuss whether these regions are actually normal or
superconducting in the following subsection.

Let us solve to find this post-compression state for the Monaghan field. This field
does not have a spatially constant magnitude $B$, nor does any self-consistent
equilibrium solution, making it less obvious which quantity should be
equated with $H_c$ to determine when compression must cease. Here, we will simply
assume this occurs when
\beq
\bar{B}
\equiv \sqrt{\frac{1}{V_{\rm inner}}\int\limits_{V_{\rm inner}}B^2\ dV}
 = H_c.
\eeq
We combine this with equation \eqref{amplify_B} to find the final radius of
the $B\neq 0$ inner region:
\beq
R_{\rm in}=\mathcal{R}\sqrt{\frac{\bar{B}_1}{H_c}}
\eeq
where $\bar{B}_1$ is the average field strength of the inner region
$r<\mathcal{R}$ at the onset of superconductivity. A similar
calculation can be done for the outer normal region, but in this case the
radius is not necessarily constrained by flux conservation, since 
the outer $0.1R_*$ of the radius of the star will be normal in
any case. So the inner radius of the outer normal region is
\beq
R_{\rm out}=\min\bigg\{0.9R_*, R_*-(R_*-\mathcal{R})\sqrt{\frac{\bar{B}_1}{H_c}}\bigg\}.
\eeq

\begin{figure}
\begin{center}
\begin{minipage}[c]{0.5\linewidth}
\includegraphics[width=\linewidth]{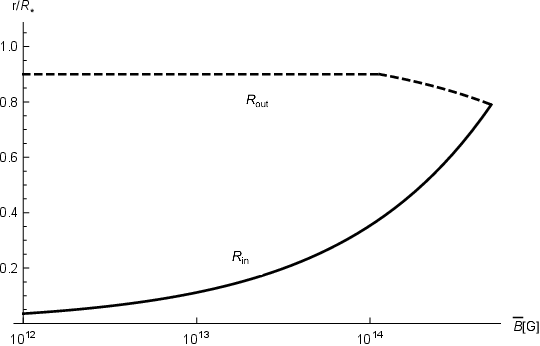}
\end{minipage}
\caption{\label{R_in_out}
Inner and outer radii of the final superconducting shell, as a
function of field strength.}
\end{center}
\end{figure}

\begin{figure}
\begin{center}
\begin{minipage}[c]{\linewidth}
\includegraphics[width=0.32\linewidth]{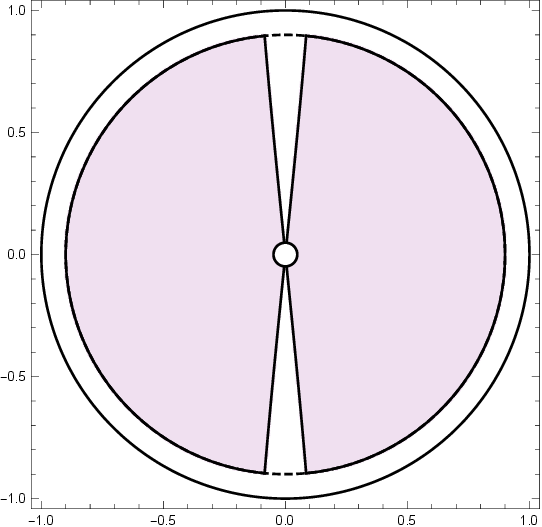}
\hspace{0.02\linewidth}
\includegraphics[width=0.32\linewidth]{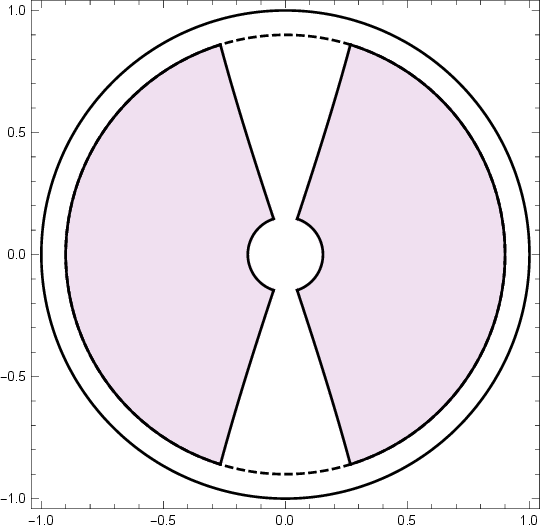}
\hspace{0.02\linewidth}
\includegraphics[width=0.32\linewidth]{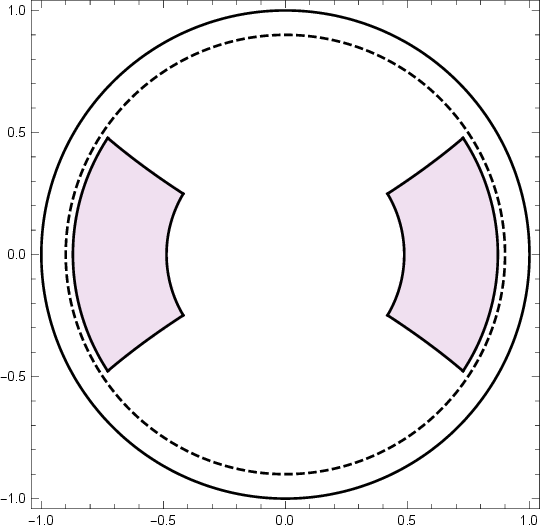}
\end{minipage}
\caption{\label{polar_hole_final}
The final structure of the superconducting region, after flux
compression has finished, for scenario 2. Scenario 1ii has an
identical structure in this 2D plane but rotated through $\pi/2$,
so that the Northern and
Southern polar `holes' in the Meissner shell then become a
single equatorial ring where the Meissner effect is not present;
scenario 1i only has the central region with magnetic field and no
breaks in the shell. For
this end state, the average field strength of
the inner region will be $H_c=5\times 10^{14}$ G. Note that whilst the
crust (outside the dashed circle but within the solid circle) is
always composed of normal matter, the white core regions could either
be normal matter or superconducting matter threaded by field lines
(the minimum-energy state for stronger fields than those we consider).}
\end{center}
\end{figure}

As well as this inner core, we note that the crustal field is also somewhat increased by the flux being
pushed outwards, by a factor of $4.4$ for the Monaghan field and our
fiducial model for superconductivity; only in the case of
$B_p=10^{14}$ G does this outer region exceed the critical field and lead to the
outer boundary $R_\mathrm{out}$ of the superconducting region being a little in from
the crust-core boundary, as seen in the right-hand panel of figure
\ref{polar_hole_final}. The variation of
$R_\mathrm{in},R_\mathrm{out}$ with average internal field strength is
plotted in figure \ref{R_in_out}.
This completes the description of the final state for scenario 1i.

For another of the four scenarios, scenario 3, there is no
calculation to perform: we assume flux threads the entire star on a
macroscopic scale, and so flux conservation does not impose any
restrictions on the final state of the magnetic field. This leaves
scenarios 1ii and 2. In both these cases $R_\mathrm{in}$ and
$R_\mathrm{out}$ are the same as for scenario 1i, but we also need to
calculate the spatial extent of the polar holes/equatorial
ring. Because the angles $\theta_\mathrm{open},\theta_\mathrm{eq}$
depend on the magnetic field, which in turn is amplified by flux
compression, we cannot just use a calculation based on the radius of
the nascent superconducting shell $\mathcal{R}$; the angular extent of
these $B\neq 0$ regions will change as the shell expands. In practice,
these can be calculated by replacing $\mathcal{R}$ in equations
\eqref{theta_open_Mon} and \eqref{theta_eq_Mon} with the changing
$R_\mathrm{in}$ and $R_\mathrm{out}$, to build up a full 2D meridional
cut through the star of the $B=0$ region at the end of the
flux-compression phase. This region is shown in purple in figure
\ref{polar_hole_final} for three different field strengths within the range where
Meissner expulsion is possible, with polar-cap values
$B_\mathrm{p}=10^{12},10^{13},10^{14}\,\mathrm{G}$, for the case of $B\neq
0$ polar holes, and so completes the description of the end state for
scenario 2. The change in opening angle throughout the shell is subtle
but noticeable (the boundaries of the holes are visibly slightly
curved). Scenario 1ii, the equatorial ring, is identical in the 2D plot except that the
plots need to be rotated by an angle of $\pi/2$.

\subsection{Are $B\neq 0$ core regions normal or superconducting?}
\label{Bneq0_sc_norm}

Throughout this paper, we have avoided discussing the state of the
core region inside the $B=0$ superconducting shell (for scenarios 1i,
1ii, and 2), and the $B\neq 0$ regions within that shell (for
scenarios 1ii and 2) - specifically, whether these $B\neq 0$ parts of
the stellar core are in a normal or superconducting state. This issue
is less important within the narrow focus of the current work, and by
side-stepping it we avoid a potentially involved discussion about the
differences between flux structures in type-I and type-II
superconductors, and normal matter. The same issue also affects the
nature of the core in scenario 3. The intention is to confront this
issue alongside the general $B\sim H_c$ case, in a follow-up
paper. Here we simply assume flux compression results in a final state
where $B=H_c$ in these regions.

The state of these $B\neq 0$ core regions is, however,
potentially crucial for related issues, such as the interactions
with the superfluid neutrons in these regions \citep{meissner_letter},
so warrants at least some brief speculation. The state is likely
determined by evolutionary processes, and understanding it is probably
inhibited by our simplistic assumption that $H_c=\mathrm{const}$ and
that we can perform calculations using an average value for the
magnetic field $\bar{B}$.

More realistically, the inner region
$r<R_\mathrm{in}$ starts normal with a field strength that is likely
to peak near/at the centre of the star and decrease away from it. Flux
compression will halt further encroachment of the superconducting
shell into this region once the outer part has reached -- probably --
a lower critical field, $H_{cI},H_{c1}$ (type-I,II, respectively). The
outer part of the $r<R_\mathrm{in}$ core would then be
superconducting, but instead of having $B=0$ would be in the
intermediate state, where flux penetrates through small-scale flux
structures. In the very central region, however, flux compression may
have taken $B$ above the upper critical field by that point -- in
which case the matter will remain normal (it does not `break'
superconductivity, as it never entered that state).

If flux compression has this effect, then it would imply that the
$B\neq 0$ holes in the superconducting shell would also be
superconducting and in the intermediate state where flux penetration
is possible -- i.e. the entire shell is superconducting, but with just
one or two contiguous regions threaded by field lines.

\section{Toroidal component and higher multipoles}

\subsection{Toroidal and linked poloidal-toroidal fields}
\label{tor_mix_fields}

We have argued that the key early phases of field rearrangement depend
on the magnetic field component that crosses the nascent shell of
superconductivity. As long as the star is approximately spherical,
$B_\perp\approx B_r$. A toroidal field is, by definition,
perpendicular to the radial direction -- so would not evolve through any scenario
involving fluid motion, making its later evolution simpler. Either it would simply be enveloped by the
expanding superconducting shell, i.e. scenario 3 in our terminology,
or could be `cut' across the incipient superconducting shell without
any need for rearrangement through advection (since toroidal field
lines are tangential to spherical shells, including the incipient
superconducting shell). In this latter case, however, matching the
field to zero across a boundary would require a surface current whose
magnitude would scale with the field strength.

\begin{figure}
\begin{center}
\begin{minipage}[c]{0.6\linewidth}
\includegraphics[width=\linewidth]{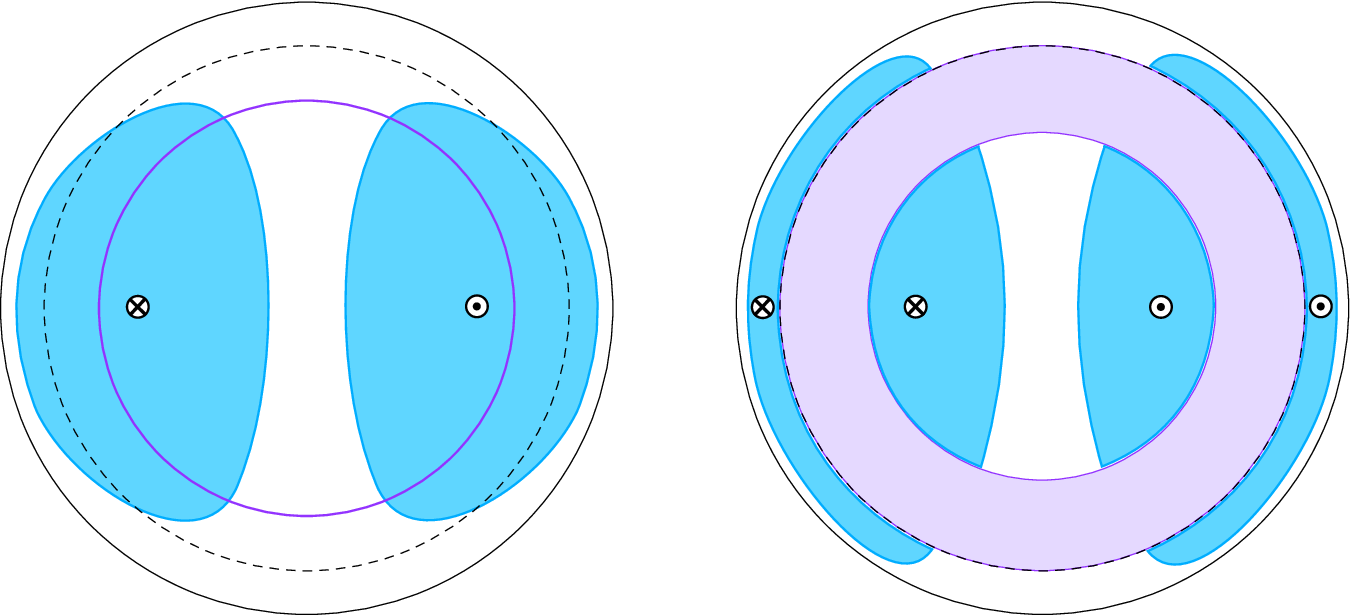}
\end{minipage}
\caption{\label{pure_tor}
A purely toroidal field has field lines everywhere perpendicular to
poloidal lines, and so orthogonal to the page in this meridional plot,
and is non-zero within some region like the blue one plotted
here. At the onset of superconductivity (left), the toroidal field on
either side of the incipient superconducting shell can be matched to
zero by having a surface current at both inner and outer
boundaries. As the shell expands (right), a complete Meissner-expelled
shell can be formed without any additional dynamics other than
compression of flux in the crust, and in the inner core region.}
\end{center}
\end{figure}

Its later evolution, including field amplification in the inner core
and crustal regions, would follow that of scenario 1i, as discussed in
section \ref{later_evol} and plotted in figure \ref{R_in_out}. There is no reason for scenarios 1ii and 2 to be
realised, i.e., if there is a Meissner-expelled region, it is likely
to be a complete shell, without $B\neq 0$ holes. The effect of the
expanding superconducting shell on a purely toroidal field, in
scenario 1i, is sketched in figure \ref{pure_tor}.

Which field configuration is most likely? Purely poloidal and purely toroidal magnetic fields are both
unrealistic, both because there is no plausible formation scenario for
these, and also because they are unstable on dynamical timescales
\citep{wright73,tayler73}, so we now consider linked
poloidal-toroidal fields where the poloidal component is at least
`significant' (not necessarily dominant).

\begin{figure}
\begin{center}
\begin{minipage}[c]{\linewidth}
\includegraphics[width=\linewidth]{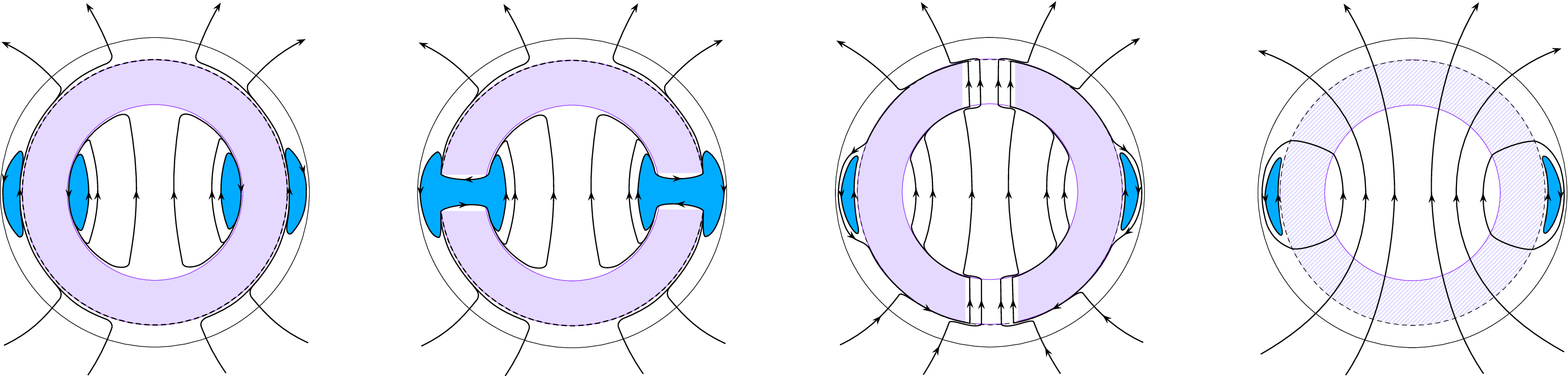}
\end{minipage}
\caption{\label{toroidal_comp}
Sketch of location of toroidal component in each scenario. Left to
right: 1i, 1ii, 2, 3.}
\end{center}
\end{figure}

The requirement $\div\bB=0$ already reduces $\bB$ to having two
degrees of freedom, poloidal and toroidal, and axisymmetry further
reduces this to one. In particular, the poloidal streamfunction $u$
becomes the single variable, and the toroidal field is expressed as a
function of $u$. This means that the toroidal component takes a
constant value along any contour of $u$, i.e. a poloidal field
line. The total extent of the region with a toroidal component is usually
chosen to be demarcated by the largest field line that closes within the
star, to avoid having the current distribution extending outside the
star \citep{LJ09}. As seen earlier (recall figure \ref{flux_scenarios}), the superconducting transition results in
various possible scenarios for the rearrangement of poloidal field
lines. In the closed-field-line region, when a toroidal component is added, these closed loops in the
plane of the page become spiral-shaped field lines that extend into the azimuthal
direction, and any fluid flow driving the rearrangement will advect
these complete field lines, not just their poloidal or toroidal
components (the latter case would also require surface currents to allow for abrupt
jumps across the normal-superconducting boundaries). Therefore, the toroidal
component must adjust to the rearranging closed poloidal field lines:
one can imagine simply `filling in' this region with toroidal
field. Doing so, we see that for scenario 1i we will be left
with two disconnected regions of toroidal field, on the inside and outside
of the superconducting shell. In scenario 1ii these regions will be
joined across the equatorial $B\neq 0$ belt to leave a region with a dumbbell-shaped
cross-section where the toroidal field resides. In both scenarios 2 and 3 the
details of the closed-field-line rearrangement depend on the exact
pre-condensation field geometry and the location of the incipient
shell of superconductivity, but we believe the mostly likely result is
a small region of toroidal field confined to the crust alone. All
these possibilities are sketched in figure \ref{toroidal_comp}.
Note the similarity of scenario 3, where poloidal fields are all open
in the core and the toroidal field is confined to the crust, to earlier equilibrium models for a
magnetic field threading the entire stellar interior and $B<H_c$ \citep{L14}.

This more realistic geometry deserves more attention, however, as the toroidal
component is not simply passive during the rearrangement, but will
play at least two roles itself. One is that the stronger the toroidal
component, the more it acts to shrink its host closed poloidal-field
region \citep{LJ09,armaza} -- thus, a very strong pre-condensation toroidal
component is more likely to be confined to a radius $r>\mathcal{R}$, which
-- following the onset of superconductivity -- will be pushed
outwards, so that the only toroidal field remaining at the end of the
Meissner phase is in the crust. The other
role the toroidal field will play is through its contribution to the overall
$B^2$, thus exceeding the critical field more easily in the
closed-field-line region and invalidating our earlier approximation that
$B\approx \mathrm{const}$ along $r=\mathcal{R}$.

\subsection{Higher-multipole pre-condensation magnetic fields}

If the magnetic field is dominated by higher multipoles, unlike the
dipolar configuration we have assumed, the characteristic lengthscale
of the field will become correspondingly shorter, affecting the likely
final state of the magnetic field. Let us assume for definiteness a
spherical harmonic expansion of the field, with the largest-amplitude
contribution being some particular harmonic $Y_l^m$. The
characteristic angular scale $\mathcal{L}_\parallel$ for the field will be reduced from
$\mathcal{L}_\parallel\sim\pi$ for the dipole field to
$\mathcal{L}_\parallel\sim\pi/l$ for the $(l,m)$ harmonic (recall that
$|m|\leq l$, so it is $\pi/l$ and not $\pi/m$ that gives the minimum
angular scale).

For the higher-multipole field, the distance one must stretch a field
line until it encounters an oppositely-directed neighbouring field
line is reduced from $\sim\pi/2$ (recall the cartoon of figure
\ref{flux_scenarios}) to $\sim\pi/(2l)$. Progressively less fluid motion is,
therefore, required to achieve this as the multipolar index
increases; in this sense it becomes `easier' to realise scenario
1i for full Meissner expulsion. The efficacy of reconnection is, however, unaffected, since
it depends upon the thickness of the reconnection zone (i.e. its
radial extent, in this context) $\mathcal{L}_\perp$ rather than $\mathcal{L}_\parallel$.

Scenarios 1ii and 2 will have a more complex structure for a
higher-multipole field; instead of being pierced once or twice, as for
the dipole field, we anticipate that the number of normal `holes' in the
superconducting shell will scale with $l$ and $m$. The higher the
multipolar index, the more likely it is that the star will end up in
scenario 3 (which can be thought of as the limiting case of scenario
1ii or 2 in the limit of very large $l,m$).

\section{Discussion and outlook}

One might find the Meissner-effect scenario proposed here implausible,
relying as it does on an interplay of two pieces of physics --
field-line advection and reconnection -- whose modelling is difficult
and only outlined here. But the motivation for this paper was not a
frivolous survey of contrived scenarios for this phase, but rather to ask the
question: is there \emph{any} way to produce a Meissner-expelled
region on a short timescale? If the answer had been no, we would have
had to conclude that essentially all numerical work on magnetic-field
evolution in the crust --
a field of research built up over the last two decades
(e.g. \citet{hollrued02,ponsgepp07,perna_pons,ascenzi}) -- would have 
been suspect, and magnetar modelling would potentially need
substantial revision (without rapid crustal evolution, one would need
to return to earlier ideas of core evolution involvement, but the
latter process is likely to be even slower). On the other hand, had
our conclusion been that the dynamical-timescale Meissner expulsion
explored here is always effective, theoretical modelling of
neutron-star rotation would have
faced challenges \citep{meissner_letter}. The model proposed here allows for both extremes, as
well as intermediate scenarios where a Meissner-expelled shell is
broken by one or two regions through which field
penetrates.

Even if more detailed future work were to show the Meissner mechanism
proposed here to be ineffective, much of our discussion about
the necessary conditions for macroscopic field expulsion would remain relevant. For
example, Ohmic decay alone reduces magnetic energy but does not
obviously have any way to induce the kind of field rearrangement needed to produce a zero-field
region. If, therefore, one wishes to invoke Ohmic decay as a mechanism for flux
transport and Meissner expulsion (as in earlier studies), one
must still confront some of the same issues discussed in this
  paper. In particular, the solenoidal constraint on the
magnetic field will again restrict how the field can be expelled,
  as will (on appropriate
timescales/lengthscales) flux conservation.

Previous work on the Meissner effect has focussed on local dynamics,
whether a small region is uniformly penetrated by flux (the normal
state), irregularly penetrated (by thin fluxtubes whose dimensions are
universal, in the case of type-II superconductivity; or other flux
structures, in the case of type-I superconductivity), or fully
expelled \citep{ho17}. The problem is particularly rich when the
proton superconductor coexists with a neutron superfluid \citep{hab_schm,wood_grab}, as
expected after a few hundred years (beyond the phase of primary
interest to us here). But
we argue here that additional restrictions need to be 
considered on the macroscopic scale, related to the field geometry, and only then can one predict
whether a given flux-transport mechanism could produce a
field-expelled region. These also lead to restrictions on the minimum
field strength for which we expect Meissner expulsion: both the
field-line advection phase and the reconnection phase require the same
value, $B\gtrsim 10^{12}\,\mathrm{G}$. This means that although, in
principle, one might expect partial Meissner expulsion for lower field
strengths than full expulsion (in the case of ineffective
reconnection, scenarios 1ii and 2), in practice this is not
realised. However, we emphasise that $B\gtrsim 10^{12}\,\mathrm{G}$
does not \emph{imply} the star will host a Meissner-expelled region,
only that it is \emph{possible}. A closer examination of the likelihood of
each of the two stages in our Meissner model (advection and
reconnection) will help to assess how
plausible each of the scenarios is, in practice. In contrast with the lower limit on $B$, the
upper limit is not given by macroscopic arguments, but rather is the same as
that for local calculations, the critical field strength. However, our
macroscopic calculations for a Meissner-expelled region with a $B\neq
0$ hole do also naturally reach the limit of
no expelled field in the case of $B\approx H_c$. 
Once the field strength reaches $H_c$ -- or if it is already globally this
strong before the onset of superconductivity -- the problem becomes
more subtle. There is no longer a single clear minimum-energy state of
$B=0$, and the resulting configuration will be an interplay between
whatever the new minimum-energy state is -- be it fluxtubes, lamellae,
or other flux structures -- and the advection and reconnection physics
needed to realise this state. A realistic model of the NS core would,
in fact, likely involve an inner core region of type-I superconductivity,
a type-II superconducting outer core, a normal crust, and suitable
conditions to describe the interface between each of these. We intend
to return to this $B\sim H_c$ case in a future study.

There are reassuring similarities between our scenario 3 and previous work on
$B<H_c$ equilibrium magnetic-field configurations in a type-II superconductor 
\citep{roberts81,hen_wass,L14}, showing that in this limit of our
modelling, the field naturally leads to the development of a global
hydromagnetic equilibrium. In cases with partial or complete
Meissner expulsion, the expected configurations are qualitatively
different. Scenario 1i features a disconnected inner region of
poloidal and toroidal field, and in scenario 1ii a $B\neq 0$
equatorial band through the field-expelled shell potentially allows
for penetration of the toroidal component into the core even in the
case of relatively weak $B$; a core toroidal component was previously
found not to occur below
$\sim 10^{14}\,\mathrm{G}$ \citep{L13,L14}. For
scenario 2, on the other hand, even in the case
$B>10^{14}\,\mathrm{G}$ the toroidal component is likely to be
confined to the crust. Finally, we showed that a stronger
pre-condensation toroidal field is more likely to be confined to the
crust post-condensation.

The model presented here invokes an unmodelled fluid flow to rearrange
magnetic field lines prior to reconnection, and the nature of this
fluid flow deserves detailed scrutiny in the future. Some residual
dynamics from birth are to be expected, but whether this is the
leading mechanism driving field-line advection remains to be
seen. What is clear is that the axisymmetric model
considered here is likely to be overly simplistic, since a
near-incompressible flow $\div\bv\approx 0$ moving consistently towards (say) the
equator in one meridional plane would have to travel some distance in
the azimuthal direction and then circulate away again. This simplistic
flow, together with the assumed initially dipolar field, make the
realisation of Meissner state in the sketches here seem rather contrived and
unlikely. However, this is just a limiting case that is conceptually
simpler. In reality, the more multipolar and small-scale the
pre-condensation field and fluid flow are,
the easier it will be to produce partial or complete Meissner
expulsion. Nonetheless, some features of the simple model, including
the restriction on the range of internal field strengths and the need
for an inner $B\neq 0$ core of matter underneath the Meissner shell,
are expected to be robust.

Some of the ideas explored here have cross-over with similar notions from stellar
dynamos. In both cases, one is concerned with understanding how a
suitable MHD flow can be rearranged and reconnected into a new
configuration, and a detailed treatment inevitably involves
assumptions about the nature of the small-scale field to be
reconnected \citep{rincon}. Both the dynamo and the Meissner-expulsion phase
of a neutron star are further complicated by the high
magnetic Prandtl number of the flow \citep{L21}. Another interesting,
related problem from fluid dynamics that may have relevance to
understanding the advection phase of our Meissner model is that of the expulsion of
magnetic flux by an eddy \citep{weiss66,gallo_weiss}.

This paper focusses on development of the theory of a dynamical
Meissner effect. A companion letter, \citet{meissner_letter}, explores
some of the interesting observational consequences of this model.


\section*{Acknowledgements}

I thank Danai Antonopoulou, Vanessa Graber, Kostas Gourgouliatos, and
Zorawar Wadiasingh for helpful discussions on some aspects of this work.

\section*{Data availability}

The data underlying this article will be made available upon reasonable request.

\bibliographystyle{mnras}

\bibliography{references}

\label{lastpage}

\end{document}